\documentclass[superscriptaddress,twocolumn,showpacs,preprintnumbers]{revtex4}
\usepackage{eurosym}
\usepackage[tbtags]{amsmath}
\usepackage{amsfonts}
\usepackage{mathrsfs}
\usepackage{amssymb}
\usepackage{graphicx}
\usepackage{epsfig,graphicx,times}
\usepackage{subfigure}
\usepackage{color}
\usepackage{amssymb}
\usepackage{amsmath}
\usepackage{graphicx}
\usepackage{dcolumn}
\usepackage{bm}
\usepackage{float}
\usepackage[abs]{overpic}
\usepackage{epstopdf}

\input{tcilatex}

\begin{document}

\title{Efficient single-photon frequency conversion in the microwave domain using superconducting quantum circuits}
\author{W. Z. Jia}
\email{wenzjia@swjtu.edu.cn}
\affiliation{School of Physical Science and Technology,
Southwest Jiaotong University, Chengdu 610031, China}
\author{Y. W. Wang}
\affiliation{School of Physical Science and Technology,
Southwest Jiaotong University, Chengdu 610031, China}
\author{Yu-xi Liu}
\email{yuxiliu@mail.tsinghua.edu.cn}
\affiliation{Institute of Microelectronics, Tsinghua University, Beijing 100084, China}
\affiliation{Tsinghua National Laboratory for Information Science and Technology (TNList), Beijing 100084, China}
\date{\today }

\begin{abstract}
We present an approach to achieve efficient single-photon frequency conversion in the microwave domain based on coherent control in superconducting quantum circuits, which consist of a driven artificial atom coupled to a semi-infinite transmission line. Using full quantum mechanical method, we analyze the single-photon scattering process in this system and find that single-photon frequency up- or down-conversion with efficiency close to unity can be achieved by adjusting the parameters of the control field applied to the artificial atom. We further show that our approach is experimentally feasible in currently available superconducting flux qubit circuits.
\end{abstract}

\maketitle

\bigskip




\bigskip

\section{\label{introduction}Introduction}

Quantum networks may be made up of hybrid quantum structures combining the advantages of different quantum systems at different energy scales~\cite{Xiang2013RMP}, where photons are usually used as ``flying'' qubits for efficient transmission of quantum information over large distances~\cite{Gisin2003Nature}, and trapped atoms (or ions, quantum dots, etc.) as ``stationary'' qubits for manipulation and storage~\cite{Kimble2007PRL,Wineland2008Nature,Yamamoto2012Nature}.  To couple different quantum systems, efficient frequency conversion of photons is required and plays important role ranging from quantum communication to quantum-information processing \cite{Kumar1990OL,Kumar1992PRL,QFC1,QFC2,QFC3,QFC4,QFC5,QFC6,QFC7}. Moreover, this technique could facilitate reliable detection of single-photon by converting the frequency of photons to a spectrum range available for photon detectors~\cite{Detec1,Detec2,Detec3}. Frequency conversion can also be used to generate photons at frequencies for which we have no suitable photon sources~\cite{ Fejer1994PhysToday}.

Recent advances in nanoscale device fabrication enable the control of light-matter interactions in ultra low power
regimes. By strongly coupling photons in nano-structures, unprecedented optical devices at single-photon 
level~\cite{Imamoglu2000Science,Gisin2008Nature,CavityQED1,CavityQED2,Shen2005OL,Shen2005PRL, 
Shen2007PRL,Shen2009PRA, Fan2010PRA,Shen2015PRA,Chang2006PRL,Chang2007NatPhys,Soresen2010NJP,Zheng2010PRA,Roy2010PRB,Roy2016} 
can be realized. Typically,  the strong light-matter interactions can be achieved by placing a quantum emitter 
inside a cavity~\cite{CavityQED1,CavityQED2}, called cavity electrodynamics (cavity-QED), or coupling a quantum emitter to a one-dimensional waveguide,  called waveguide electrodynamics (waveguide-QED)~\cite{Shen2005OL,Shen2005PRL, Shen2007PRL,Shen2009PRA,Fan2010PRA,Shen2015PRA,Chang2006PRL,Chang2007NatPhys,Soresen2010NJP,Zheng2010PRA, 
Roy2010PRB, Roy2016}. Based on waveguide-QED structure, some effective single-photon frequency conversion 
schemes~\cite{Shen2012PRL, Shen2012PRA, Wang2014PRA} have been proposed.

The waveguide-QED structure can be realized based on superconducting circuits by coupling a superconducting artificial atom to a one-dimensional transmission line ~\cite{Tsai2010Science, Tsai2010PRL1, Tsai2010PRL2, Delsing2011PRL, Delsing2012PRL, Delsing2013PRL, Delsing2015NATPHYS}. We know that the operating frequency for waveguide QED system is usually determined by the material properties of the quantum emitter and can not be engineered. However, for superconducting artificial atoms, e.g., flux qubit~\cite{fluxqubit}, fluxonium~\cite{fluxonium}, and phase qubits~\cite{phasequbit}, the spacings of energy levels and transition elements are tunable by adjusting external variables~\cite{DeltaAA1, DeltaAA2, DeltaAA3}. Meanwhile, recent experiments show that these systems exhibit high atom-waveguide coupling efficiency~\cite{Tsai2010Science, Tsai2010PRL1, Tsai2010PRL2, Delsing2011PRL, Delsing2012PRL, Delsing2013PRL, Delsing2015NATPHYS}, meaning low leakage of photons into nonwaveguided degrees of freedom. In this paper, by utilizing these advantages of superconducting quantum circuits, we propose an approach to realize efficient frequency conversion for microwave single-photons with tunable input and output frequency.  Different from existing methods using a Josephson parametric converter~\cite{Devoret2013PRL}, or using dressed-state engineering of a driven circuit-QED system~\cite{Nakamura2014PRL}, our proposal is based on a waveguide-QED structure that a driven three level artificial atom with $\Delta$-type transition~\cite{DeltaAA1, DeltaAA4, DeltaAA5, DeltaAA6, DeltaAA7} (e.g., flux qubit) is coupled to a semi-infinite transmission line. Using a full quantum mechanical method, we find that by adjusting the parameters of the classical control field, the proposed device can achieve single-photon frequency up- or down-conversion with efficiency close to unity. Note that existing proposals of single-photon frequency conversion based on waveguide-QED structures~\cite{Shen2012PRL,Shen2012PRA,Nakamura2014PRL,Zueco2016PRA}, to obtain a high conversion efficiency, the ratio between different decay rates of the emitter should satisfy particular condition. However, our approach, benefiting from the control field, allows arbitrary decay rates of the transition channels relevant to single-photon frequency conversion. We also note that recently three-wave mixing~\cite{TWMinSQC} and frequency conversion~\cite{Zhao2015arxiv} of classical microwave fields via a single three-level superconducting qubit has been studied.

The paper is organized as follows. In Sec.~\ref{SP-fre-conv}, we give a general theoretical model 
for single-photon frequency conversion, including Hamiltonian and equations of motion in Sec.~\ref{model}, the scattering problem of a monochromatic single-photon in Sec.~\ref{photon-scattering}, and the scattering problem of a single-photon with finite bandwidth and corresponding frequency-conversion efficiency in Sec.~\ref{scattering-with-BW}. Then, in Sec.~\ref{Experimental-realization}, using experimentally feasible parameters, we study a physical realization of proposed frequency convertor using a superconducting flux qubit embedded at the end of a semi-infinite one-dimensional transmission line. Finally, further discussions and conclusions are given in Sec.~\ref{Conclusions}.

\begin{figure}[t]
\centering
\subfigure[]{
\label{Fig.sub.1}
\includegraphics[width=0.275\textwidth]{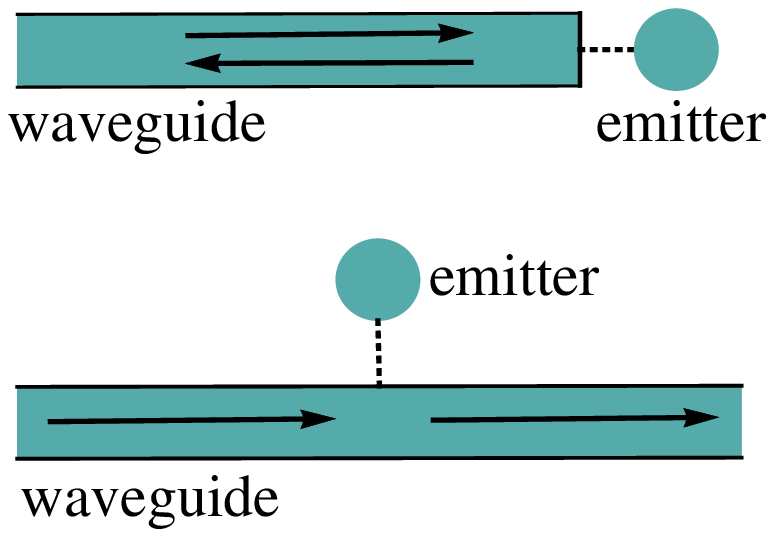}}
\subfigure[]{
\label{Fig.sub.2}
\includegraphics[width=0.175\textwidth]{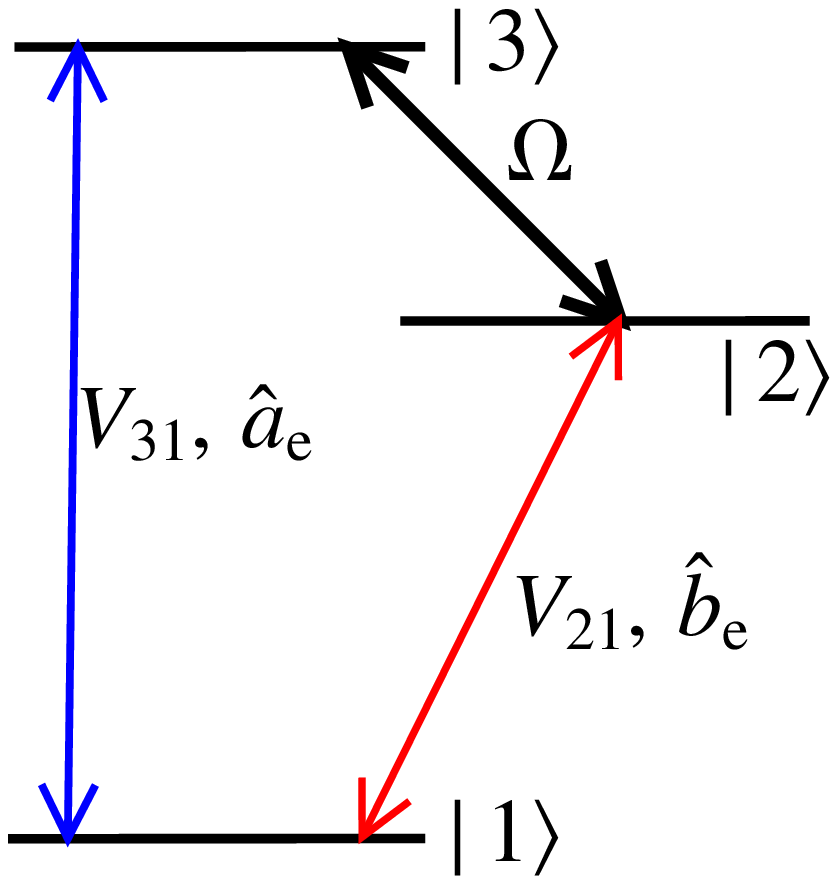}}
\caption{(Color online) (a) Schematics of a quantum emitter coupled to a one-dimensional waveguide, Œin which single 
photons propagate along the arrow direction. The upper part shows the directly-coupled cases. The lower part 
shows the side-coupled cases.  (b) Energy level structure of the $\Delta$-type emitter (artificial atom). The transitions
$|1\rangle\leftrightarrow |2\rangle$ and $|1\rangle\leftrightarrow |3\rangle$ are coupled to the waveguide modes 
with strength $V_{21}$ and $V_{31}$, respectively. The transition $|2\rangle\leftrightarrow |3\rangle$ is driven by a
classical control field with the Rabi frequency $\Omega$ƒ.}
\label{WG_ATOM}
\end{figure}

\section{\label{SP-fre-conv}Single-photon frequency conversion}
 
\subsection{\label{model}Hamiltonian and equations of motion}

In this section, we give a general theoretical framework for the frequency-conversion process. 
The system consists of a $\Delta$-type three-level quantum emitter being placed at one end of a semi-infinite 
waveguide, as shown in the upper schematic diagram in Fig.~\ref{WG_ATOM}(a). The quantum emitter can be flux qubit~\cite{fluxqubit}, fluxonium~\cite{fluxonium}, phase qubit~\cite{phasequbit}, or other types of superconducting artificial atoms, and here the ``waveguide'' represents a transmission line. This coupling configuration is widely adopted in experiments based on superconducting quantum circuits~\cite{CapacityCoupl1,CapacityCoupl2}. In this case, the propagation of the right- and left-going modes in the transmission line is then restricted to $x < 0$. They can be easily mapped to a unidirectional transporting even modes~\cite{Shen2009PRA,Soresen2010NJP}, as shown in the lower schematic diagram in Fig.~\ref{WG_ATOM}(a). 
For convenience, our calculation on single-photon scattering will be based on this equivalent configuration. Note that this type of chiral modes is important to realize a high conversion efficiency, as discussed in Refs.~\cite{Shen2012PRL,Shen2012PRA}, where a Sagnac interferometer is utilized to generate effective even modes.  The energy level structure of the quantum emitter is shown in Fig.~\ref{WG_ATOM}(b). Three energy levels are denoted by $|1\rangle$, $|2\rangle$, and  $|3\rangle$, respectively. They possess $\Delta$-type transition. In our proposal, the transition channels $|1\rangle\leftrightarrow|3\rangle$ and $|1\rangle\leftrightarrow|2\rangle$ are used to interact with single-photons relevant to frequency conversion. Thus the microwave modes coupling to these transitions are modeled as quantized fields. Moreover, a classical microwave, acting as a control field, with frequency $\omega$ and Rabi frequency $\Omega$,… is applied to couple the levels $|2\rangle$ and $|3\rangle$.

In real space, the total Hamiltonian describing the system shown in the lower schematic diagram  in Fig.~\ref{WG_ATOM}(a) can be written as
\begin{equation}
\hat{H}=\hat{H}_{\mathrm{WG}}+\hat{H}_{\mathrm{atom}}+\hat{H}_{\mathrm{d}}+\hat{H}_{\mathrm{int}},
\label{Hamiltonian}
\end{equation}
where $\hat{H}_{\mathrm{WG}}$ describes the free propagation of the photons in the waveguide, $\hat{H}_{\mathrm{atom}}$ describes the quantum emitter, $\hat{H}_{\mathrm{int}}$ describes the interaction between the quantized fields and the emitter, and $\hat{H}_{\mathrm{d}}$ describes the interaction between the classical driving field and the emitter.

The free-photon Hamiltonian can be written as
\begin{eqnarray}
\hat{H}_{\mathrm{WG}}/\hbar&=&\int \mathrm{d}x\hat{a}_{\mathrm{e}}^{\dagger}\left(x\right)\left(-\mathrm{i} v_{\mathrm{g}}\frac{\partial}{\partial
x}\right)\hat{a}_{\mathrm{e}}\left(x\right)
\nonumber
\\
&&+\int \mathrm{d}x\hat{b}_{\mathrm{e}}^{\dagger}\left(x\right)\left(-\mathrm{i} v_{\mathrm{g}}\frac{\partial}{\partial
x}\right)\hat{b}_{\mathrm{e}}\left(x\right),
\label{H_WG}
\end{eqnarray}
where $v_{\mathrm{g}}$ is the group velocity of the photons. 
Here we have assumed that the entire frequency range of interest is far away from the cutoff frequency of the waveguide, so that the linear dispersion relation holds. In addition, we assume the transitions frequencies greatly exceeds the linewidths (given by the effective decay rate).  As a result, only modes in a very narrow frequency interval around the energy levels  can efficiently interact with the emitter. Thus we can safely treat the waveguide modes as two distinct ones, i.e., $a$- and $b$-modes. Specifically, in Hamiltonian~\eqref{H_WG}, $\hat{a}_{\mathrm{e}}^{\dagger}(x)$~[$\hat{a}_{\mathrm{e}}(x)$] is bosonic operators creating(annihilating) an $a$-mode photon at $x$, and $\hat{b}_{\mathrm{e}}^{\dagger}(x)$~[$\hat{b}_{\mathrm{e}}(x)$] is the bosonic operator of $b$-mode photons. Note that both $a$- and $b$-modes are even modes~\cite{Shen2009PRA,Soresen2010NJP}, thus we use the subscript ``e'' to denotes them.
 
The atomic Hamiltonian is given by
\begin{equation}
\hat{H}_{\mathrm{atom}}/\hbar =
\sum_{i=2,3}\left(\omega_{i1}-\mathrm{i}\frac{\gamma_{i}}{2}\right)\left\vert i\right\rangle \left\langle i\right\vert,
\label{H_atom}
\end{equation}
where $\omega_{i1}$ is the transition frequency from the state $\vert i\rangle$ to the state $\vert 1\rangle$. Here the energy of the ground state $|1\rangle$ is set 
to zero as reference. Additionally, in the spirit of the quantum jump picture~\cite{CarmichaelBook},  we introduce imaginary parts 
$-i\gamma_{i}/2$ in the Hamiltonian. These dissipation terms model the pure dephasing due to 
fluctuations of atomic levels and the photon loss due to
coupling to the modes other than waveguide  $a$- and $b$-modes.  

Under rotating wave approximation, the interaction between the classical driving field and the emitter can be represented by the following Hamiltonian : 
\begin{equation}
\hat{H}_{\mathrm{d}}/\hbar=-\frac{\Omega}{2}\left(\left\vert 3\right\rangle \left\langle 2\right\vert \mathrm{e}^{-\mathrm{i}\omega t}+\mathrm{H.c.}\right),
\label{H_d}
\end{equation}
where $\omega$ and $\Omega$ are the frequency and Rabi frequency of the classical field, respectively. Without loss of generality, the Rabi frequency is assumed to be real.

Under rotating wave approximation, the interaction between the quantized modes in waveguide and the emitter is represented by the following Hamiltonian :
\begin{eqnarray}
&&\hat{H}_{\mathrm{int}}/\hbar = -V_{31}\int
\mathrm{d}x\delta\left(x\right)\left[\hat{a}_{\mathrm{e}}^{\dagger}\left(x\right)\left\vert 1\right\rangle \left\langle 3\right\vert+\mathrm{H.c.}\right]
\nonumber
\\
&& -V_{21}\int
\mathrm{d}x\delta\left(x\right)\left[\hat{b}_{\mathrm{e}}^{\dagger}\left(x\right)\left\vert 1\right\rangle \left\langle 2\right\vert+\mathrm{H.c.}\right],
\label{H_int}
\end{eqnarray}
where $V_{21}$ and $V_{31}$ are the coupling strengths. 

For single-photon scattering problem, a general interaction state $\left\vert\Psi\left(t\right)\right\rangle$ can be expanded in the single-photon subspace as
\begin{eqnarray}
&&\left\vert\Psi\left(t\right)\right\rangle=\int{\mathrm{d}x}\phi_{a}\left(x,t\right)\hat{a}_{\mathrm{e}}^{\dagger}\left(x\right)\left\vert \varnothing,1\right\rangle
\nonumber
\\
&&+\int{\mathrm{d}x}\phi_{b}\left(x,t\right)\hat{b}_{\mathrm{e}}^{\dagger}\left(x\right)\left\vert \varnothing,1\right\rangle
+\sum_{i=2,3}{\lambda_{i}\left(t\right)\left\vert \varnothing,i\right\rangle},
\label{statevec}
\end{eqnarray}
where $\left\vert \varnothing,1\right\rangle$ is the vacuum state, representing zero photon
in the waveguide and the atom in
its ground state. $\left\vert \varnothing,i\right\rangle$ is the $0$-photon state with the atom in the exited state $\left\vert i\right\rangle$~($i=2,3$). $\phi_{a}\left(x,t\right)$~[$\phi_{b}\left(x,t\right)$] is the wave 
function of $a$-($b$-) mode single-photons. $\lambda_{i}(t)$ is excitation amplitude of the
the atomic level $| i\rangle$. The dynamics of system is governed by the  Schr\"odinger equation
\begin{equation}
\hat{H}\left\vert\Psi\left(t\right)\right\rangle=\mathrm{i}\hbar\frac{\partial}{\partial
t}\left\vert\Psi\left(t\right)\right\rangle.
\label{SchrodingerEQ}
\end{equation}

By substituting Eq.~\eqref{statevec} into Eq.~\eqref{SchrodingerEQ}, we obtain the following equations of motion:
\begin{eqnarray}
&&-\mathrm{i} v_{\mathrm{g}}\frac{\partial}{\partial{x}}\phi_{a}\left(x,t\right)-V_{31}\delta\left(x\right)
\lambda_{3}\left(t\right)=\mathrm{i}\frac{\partial{\phi_{a}\left(x,t\right)}}{\partial{t}},
\label{eqns1}
\\
&&-\mathrm{i} v_{\mathrm{g}}\frac{\partial}{\partial{x}}\phi_{b}\left(x,t\right)-V_{21}\delta\left(x\right)
\lambda_{2}\left(t\right)=\mathrm{i}\frac{\partial{\phi_{b}\left(x,t\right)}}{\partial{t}},
\label{eqns2}
\\
&&-V_{21}\phi_{b}\left(0,t\right)+\left(\omega_{21}-\frac{\mathrm{i} \gamma_{2}}{2}\right)\lambda_{2}\left(t\right)
-\frac{\Omega}{2}\mathrm{e}^{\mathrm{i}\omega t}\lambda_{3}\left(t\right)
\nonumber
\\
&&=\mathrm{i}\frac{\partial{\lambda_{2}\left(t\right)}}{\partial{t}},
\label{eqns3}
\\
&&-V_{31}\phi_{a}\left(0,t\right)+\left(\omega_{31}-\frac{\mathrm{i} \gamma_{3}}{2}\right)\lambda_{3}\left(t\right)
-\frac{\Omega}{2}\mathrm{e}^{-\mathrm{i}\omega t}\lambda_{2}\left(t\right)
\nonumber
\\
&&=\mathrm{i}\frac{\partial{\lambda_{3}\left(t\right)}}{\partial{t}}.
\label{eqns4}
\end{eqnarray}

\subsection{\label{photon-scattering}Single-photon scattering spectra}

\subsubsection{\label{Down-conversion}Down-conversion}

We first deal with the frequency down-conversion problem. Let us assume that the atom is initially ($t\rightarrow-\infty$) in its ground state and a monochromatic $a$-mode photon (i.e., its frequency is near $\omega_{31}$) is incident, the corresponding state of the whole system can be written as
\begin{equation}
\vert\Psi_{\mathrm{in}}(t)\rangle=\vert\nu\rangle_{a}=\frac{1}{\sqrt{2\pi}}\int\mathrm{d}x e^{\mathrm{i} k x-\mathrm{i} {\nu} t}\hat a_{\mathrm{e}}^{\dagger}\left(x\right)\left\vert\varnothing,1\right\rangle,
\label{initial-a} 
\end{equation}
where $\nu$ and $k$ are frequency and wave vector of photon, satisfying dispersion relation $\nu=v_{\mathrm{g}} k$. Under this initial condition, when the atom interacts with the photon, the excitation amplitude $\lambda_3(t)$ in the state~\eqref{statevec} should oscillate at the incoming-photon frequency $\nu$, and the excitation amplitude $\lambda_2(t)$ should oscillate at the difference frequency $\nu'=\nu-\omega$ between the incoming photon and the 
classical field.  After scattering, the frequency of outgoing photon may either stay unchanged or experience a red 
shift. Thus the amplitudes of the state vector~\eqref{statevec} should take the following ansatz: 
\begin{eqnarray}
&&\phi_{a}(x,t)=\frac{1}{\sqrt{2\pi}}e^{\mathrm{i} k x-\mathrm{i} {\nu} t}\left[\theta(-x)+T_{a}\theta(x)\right],
\label{ansatz1}
\\
&&\phi_{b}(x,t)=\frac{1}{\sqrt{2\pi}} e^{\mathrm{i} k' x-\mathrm{i} {\nu}' t}T_{b}\theta(x),
\label{ansatz2}
\\
&&\lambda_{2}(t)=\Lambda_{2}e^{-\mathrm{i}\nu' t},
\label{ansatz3}
\\
&&\lambda_{3}(t)=\Lambda_{3}e^{-\mathrm{i}\nu t},
\label{ansatz4}
\end{eqnarray}
where $T_{a}$ and $T_{b}$ are the transmission coefficients of $a$- and $b$-mode photon,
respectively. $k'=k-\omega/v_{\mathrm{g}}$ is the wave vector of red-shifted photon, $\theta(x)$ denotes the Heaviside step function, and $\Lambda_{i}$ ($i=2,3$) is the time independent part of $\lambda_{i}(t)$. Clearly, $\left\vert T_{a}\right\vert^2$  gives
the probability of elastic scattering, ( i.e., the outgoing photon experiencing no frequency shift), and  $\left\vert T_{b}\right\vert^2$  gives the probability of inelastic scattering (i.e., a photon with down-shifted frequency $\nu'=\nu-\omega$ is generated). 

Substituting Eqs.~\eqref{ansatz1}-\eqref{ansatz4} into Eqs.~\eqref{eqns1}-\eqref{eqns4}, the equations for  transmission coefficients of photon and excitation amplitudes of emitter are given by
\begin{eqnarray}
&&- \frac{\mathrm{i}v_{\mathrm{g}}}{\sqrt{2\pi}}\left(T_{a}-1\right)-V_{31}\Lambda_{3}=0,
\label{eqncc1}
\\
&&- \frac{\mathrm{i}v_{\mathrm{g}}}{\sqrt{2\pi}}T_{b}-V_{21}\Lambda_{2}=0,
\label{eqncc2}
\\
&&-\frac{V_{21}}{2\sqrt{2\pi}}T_{b}+\left(\omega_{21}-{\nu}'-\mathrm{i}\frac{\gamma_{2}}{2}\right)\Lambda_{2}-\frac{\Omega}{2}\Lambda_{3}=0,
\label{eqncc3}
\\
&&-\frac{V_{31}}{2\sqrt{2\pi}}\left(T_{a}+1\right)+\left(\omega_{31}-{\nu}-\mathrm{i}\frac{\gamma_{3}}{2}\right)\Lambda_{3}-\frac{\Omega}{2}\Lambda_{2}=0.
\nonumber
\\
\label{eqncc4}
\end{eqnarray}
Solving Eqs.~\eqref{eqncc1}-\eqref{eqncc4}, we have

\begin{figure*}[t]
\centering
\includegraphics[width=1\textwidth]{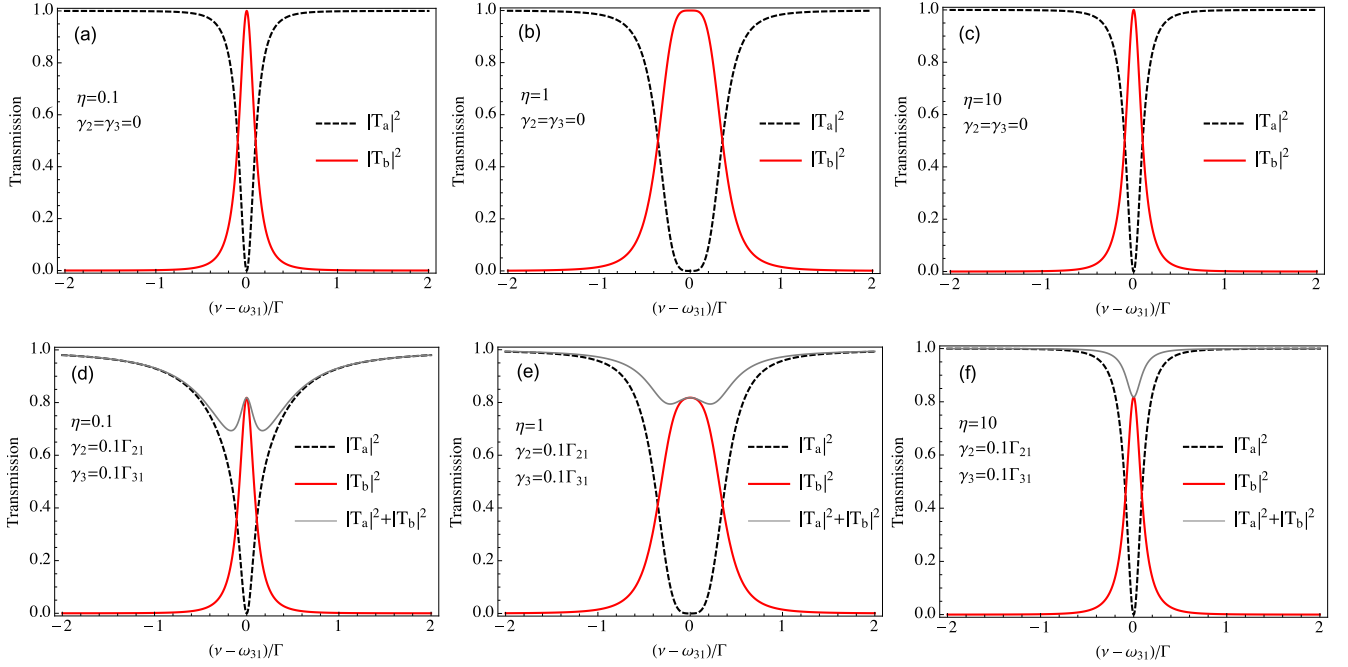}
\caption{(Color online) Transmission spectra of frequency down-conversion case. $|T_{a}|^2$ represents elastic transmission, and $|T_{b}|^2$ represents inelastic transmission.  (a)-(c) give the transmission
spectra of the ideal case with no photon loss; (d)-(f) give the transmission
spectra with photon loss.}
\label{TransmissionSpectrum}
\end{figure*}
\begin{widetext}
\begin{eqnarray}
&&T_{a}(\nu)=\frac{\left[\mathrm{i}\left(\nu-\omega_{31}\right)+\frac{\Gamma_{31}}{2}-\frac{\gamma_{3}}{2}\right]\left[\mathrm{i}\left(\nu-\omega_{31}-\Delta\right)
-\frac{\Gamma_{21}}{2}-\frac{\gamma_{2}}{2}\right]+\frac{\Omega^2}{4}}{\left[\mathrm{i}\left(\nu-\omega_{31}\right)-\frac{\Gamma_{31}}{2}-\frac{\gamma_{3}}{2}\right]\left[\mathrm{i}\left(\nu-\omega_{31}-\Delta\right)
-\frac{\Gamma_{21}}{2}-\frac{\gamma_{2}}{2}\right]+\frac{\Omega^2}{4}},
\label{TranTa}
\\
&&T_{b}(\nu)=\frac{-\frac{\mathrm{i}}{2}\sqrt{\Gamma_{21}\Gamma_{31}}\Omega}{\left[\mathrm{i}\left(\nu-\omega_{31}\right)-\frac{\Gamma_{31}}{2}-\frac{\gamma_{3}}{2}\right]\left[\mathrm{i}\left(\nu-\omega_{31}-\Delta\right)
-\frac{\Gamma_{21}}{2}-\frac{\gamma_{2}}{2}\right]+\frac{\Omega^2}{4}},
\label{TranTb}
\\
&&\Lambda_{2}(\nu)=\frac{1}{\sqrt{2\pi}}\frac{-\frac{1}{2}V_{31}\Omega}{\left[\mathrm{i}\left(\nu-\omega_{31}\right)-\frac{\Gamma_{31}}{2}-\frac{\gamma_{3}}{2}\right]\left[\mathrm{i}\left(\nu-\omega_{31}-\Delta\right)
-\frac{\Gamma_{21}}{2}-\frac{\gamma_{2}}{2}\right]+\frac{\Omega^2}{4}},
\\
&&\Lambda_{3}(\nu)=\frac{1}{\sqrt{2\pi}}\frac{-\mathrm{i} V_{31}\left[\mathrm{i}\left(\nu-\omega_{31}-\Delta\right)
-\frac{\Gamma_{21}}{2}-\frac{\gamma_{2}}{2}\right]}{\left[\mathrm{i}\left(\nu-\omega_{31}\right)-\frac{\Gamma_{31}}{2}-\frac{\gamma_{3}}{2}\right]\left[\mathrm{i}\left(\nu-\omega_{31}-\Delta\right)
-\frac{\Gamma_{21}}{2}-\frac{\gamma_{2}}{2}\right]+\frac{\Omega^2}{4}},
\label{freDC}
\end{eqnarray}
\end{widetext}
where $\Gamma_{ij}=V_{ij}^2/v_{\mathrm{g}}$ represents atom-waveguide decay rate through transition channel $|i\rangle\leftrightarrow |j\rangle$. $\Delta=\omega-(\omega_{31}-\omega_{21})$ is the detuning of the classical driving field.

We find that efficient frequency conversion can be realized by adjusting the parameters of the classical field. Specifically, if we set
\begin{equation}
\Delta=0,~~~\Omega=\sqrt{\left(\Gamma_{31}-\gamma_{3}\right)\left(\Gamma_{21}+\gamma_{2}\right)},
\label{opt-condition-DC}
\end{equation}
then the transmission probabilities for an incident photon on-resonance with the emitter (i.e., $\nu=\omega_{31}$) are 
\begin{equation}
|T_{a}(\omega_{31})|^2=0,~~~|T_{b}(\omega_{31})|^2=\frac{1-{\gamma_{3}}/{\Gamma_{31}}}{1+{\gamma_{2}}/{\Gamma_{21}}}.
\label{trans-DC}
\end{equation}
This result shows that after interacting with the emitter,  an incoming $a$-mode photon is totally converted to a photon of other modes ($b$-mode or reservoir). In particular, if there is no photon loss ($\gamma_{2}=\gamma_{3}=0$), an incident photon with frequency $\omega_{31}$ will completely convert to a photon with frequency $\omega_{31}-\omega$. In fact, if we rewrite Eqs.~\eqref{eqncc1} and \eqref{eqncc2} as 
\begin{eqnarray}
&&T_{b}=\mathrm{i}\sqrt{2\pi}\frac{V_{21}}{v_{\mathrm{g}}}\Lambda_{2},
\label{Tb}
\\
&&T_{a}=1+\mathrm{i}\sqrt{2\pi}\frac{V_{31}}{v_{\mathrm{g}}}\Lambda_{3},
\label{Ta}
\end{eqnarray}
then one can find that the elastic-scattering photon results from the interference between the directly transmitted photon [represented by the first term in Eq.~\eqref{Ta}] and the re-emitted photon by the atom [represented by the second term in Eq.~\eqref{Ta}]. Under the optimal frequency conversion condition~\eqref{opt-condition-DC},  the interference is destructive, giving zero transmission (i.e.,$T_{a}=0$) for an $a$-mode photon on-resonance with atomic transition $|1\rangle\leftrightarrow|3\rangle$. If there is no photon loss, according to photon number conservation, the incident $a$-mode single photon must convert to a $b$-mode single photon. 

Now we discuss the transmission spectra in more details. When the emitter dissipation rate is not included (i.e., $\gamma_{2}=\gamma_{3}=0$), the condition \eqref{opt-condition-DC} becomes 
\begin{equation}
\Delta=0,~~~\Omega=\sqrt{\Gamma_{31}\Gamma_{21}}.
\label{opt-condition-DCwithoutLoss}
\end{equation}
The corresponding transmission probabilities $\left\vert T_{a}(\nu)\right\vert^2$ and $\left\vert T_{b}(\nu)\right\vert^2$ are
\begin{eqnarray}
\left\vert T_{a}(\nu)\right\vert^2=\frac{\left(\eta^2-1\right)^2\left(\frac{\nu-\omega_{31}}{\Gamma}\right)^2+4\left(\eta+1\right)^4\left(\frac{\nu-\omega_{31}}{\Gamma}\right)^4}{\eta^2+\left(\eta^2-1\right)^2\left(\frac{\nu-\omega_{31}}{\Gamma}\right)^2+4\left(\eta+1\right)^4\left(\frac{\nu-\omega_{31}}{\Gamma}\right)^4},
\nonumber
\\
\\
\left\vert T_{b}(\nu)\right\vert^2=\frac{\eta^2}{\eta^2+\left(\eta^2-1\right)^2\left(\frac{\nu-\omega_{31}}{\Gamma}\right)^2+4\left(\eta+1\right)^4\left(\frac{\nu-\omega_{31}}{\Gamma}\right)^4},
\nonumber
\\
\end{eqnarray}
where $\Gamma=\Gamma_{21}+\Gamma_{31}$ and $\eta=\Gamma_{21}/\Gamma_{31}$. For given $\Gamma$, we have relation $\left\vert T_{a}(\eta)\right\vert^2=\left\vert T_{a}(1/\eta)\right\vert^2$ and $\left\vert T_{b}(\eta)\right\vert^2=\left\vert T_{b}(1/\eta)\right\vert^2$, as shown in Figs.~\ref{TransmissionSpectrum}(a)-\ref{TransmissionSpectrum}(c). For different $\eta$,
highly efficient frequency down-conversion can be realized around the resonant point $\nu=\omega_{31}$ if our driving field satisfies
condition  \eqref{opt-condition-DCwithoutLoss}, as shown in Figs.~\ref{TransmissionSpectrum}(a)-\ref{TransmissionSpectrum}(c). 
Specifically, if $\eta\sim 1$, the line width of the transmission curve is on the order of $\Gamma_{21}$ (or $\Gamma_{31}$).   
If $\eta\gg 1$,  the transmission spectra around the resonant point can be approximated as  $\left\vert T_{a}\right\vert^2\simeq{(\nu-\omega_{31})^2}/{[\Gamma_{31}^2+(\nu-\omega_{31})^2]}$ and $\left\vert T_{b}\right\vert^2\simeq{\Gamma_{31}^2}/{[\Gamma_{31}^2+(\nu-\omega_{31})^2]}$, 
exhibiting inverted Lorentzian and Lorentzian line shape, respectively,  as shown in Fig.~\ref{TransmissionSpectrum}(c). Clearly, only a photon with frequency ranging from $\omega_{31}-\Gamma_{31}$ to $\omega_{31}+\Gamma_{31}$ can be efficiently converted to a frequency down-shifted photon.  
If $\eta\ll 1$,  the transmission spectra around the resonant point can be approximated as $\left\vert T_{a}\right\vert^2\simeq{(\nu-\omega_{31})^2}/{[\Gamma_{21}^2+(\nu-\omega_{31})^2]}$ and $\left\vert T_{b}\right\vert^2\simeq{\Gamma_{21}^2}/{[\Gamma_{21}^2+(\nu-\omega_{31})^2]}$.
In this case, only a photon with frequency ranging from $\omega_{31}-\Gamma_{21}$ to $\omega_{31}+\Gamma_{21}$ can be efficiently converted to a frequency down-shifted photon,  as shown in Fig.~\ref{TransmissionSpectrum}(a). 

It should be emphasized that in our proposal of frequency-conversion,   even if the ratio between $\Gamma_{21}$ and $\Gamma_{31}$ is arbitrary, conversion efficiency close to unity can be achieved in the presence of the tunable control field $\Omega$. While in the schemes based on $\Lambda$-type quantum emitter \cite{Shen2012PRL, Shen2012PRA}, to obtain a high conversion efficiency, equal decay rates (i.e., atom-waveguide coupling rates) for different transition channels are required, which is not easy to be achieved in real cases.

In practice, there always exists photon loss due to emitter coupling to environment. 
The transmission spectra in the presence of atomic dissipation are shown in Figs.~\ref{TransmissionSpectrum}(d)-(f). If there are no dissipative processes, the sum of the transmission coefficients should satisfy $\left\vert T_{a}\right\vert^2 + \left\vert T_{b}\right\vert^2 = 1$. However, when the atom dissipation rate is included, the leakage of photons into the degrees of freedom other than
waveguide modes (i.e., $a$- and $b$-modes) can lead to $\left\vert T_{a}\right\vert^2 + \left\vert T_{b}\right\vert^2 < 1$, as shown in the gray thin curves in Figs.~\ref{TransmissionSpectrum}(d)-(f). To obtain a highly efficient frequency conversion, we need to guide most of the decayed photons into waveguide modes, i.e., 
\begin{equation}
\gamma_{3}\ll\Gamma_{31}, ~~\gamma_{2}\ll\Gamma_{21}. 
\label{DecayCondition}
\end{equation}
In this case, according to Eq.~\eqref{trans-DC}, the transmission probability of an on-ressonance $b$-mode photon can be approximated as $|T_{b}(\omega_{31})|^2\simeq1-\gamma_{2}/\Gamma_{21}-\gamma_{3}/\Gamma_{31}$ 
[see Figs.~\ref{TransmissionSpectrum}(d)-\ref{TransmissionSpectrum}(f)].   

\subsubsection{\label{Up-onversion}Up-conversion}

If the emitter is initially prepared in its ground state $|1\rangle$, the frequency of an incident $b$-mode photon (with frequency near $\omega_{21}$) can also be up-converted. In this case, the initial state of system is
\begin{equation}
\vert\Psi_{\mathrm{in}}(t)\rangle=\vert\nu\rangle_{b}=\frac{1}{\sqrt{2\pi}}\int\mathrm{d}x e^{\mathrm{i} k x-\mathrm{i} {\nu} t}\hat b_{\mathrm{e}}^{\dagger}\left(x\right)\left\vert\varnothing,1\right\rangle.
\label{initial-b} 
\end{equation}
The calculations are similar to those of frequency down-conversion, and the main results are summarized below. The transmission coefficients of photon and the excitation amplitudes of emitter are given by 
\begin{widetext}
\begin{eqnarray}
&&\tilde{T}_{a}(\nu)=\frac{-\frac{\mathrm{i}}{2}\sqrt{\Gamma_{21}\Gamma_{31}}\Omega}{\left[\mathrm{i}\left(\nu -\omega_{21}+\Delta\right)-\frac{\Gamma_{31}}{2}-\frac{\gamma_{3}}{2}\right]\left[\mathrm{i}\left(\nu -\omega_{21}\right)
-\frac{\Gamma_{21}}{2}-\frac{\gamma_{2}}{2}\right]+\frac{\Omega^2}{4}},
\\
&&\tilde{T}_{b}(\nu )=\frac{\left[\mathrm{i}\left(\nu-\omega_{21}+\Delta\right)-\frac{\Gamma_{31}}{2}-\frac{\gamma_{3}}{2}\right]\left[\mathrm{i}\left(\nu -\omega_{21}\right)
+\frac{\Gamma_{21}}{2}-\frac{\gamma_{2}}{2}\right]+\frac{\Omega^2}{4}}{\left[\mathrm{i}\left(\nu -\omega_{21}+\Delta\right)-\frac{\Gamma_{31}}{2}-\frac{\gamma_{3}}{2}\right]\left[\mathrm{i}\left(\nu -\omega_{21}\right)
-\frac{\Gamma_{21}}{2}-\frac{\gamma_{2}}{2}\right]+\frac{\Omega^2}{4}},
\\
&&\tilde{\Lambda}_{2}(\nu )=\frac{1}{\sqrt{2\pi}}\frac{-\mathrm{i} V_{21}\left[\mathrm{i}\left(\nu -\omega_{21}+\Delta\right)-\frac{\Gamma_{31}}{2}-\frac{\gamma_{3}}{2}\right]}{\left[\mathrm{i}\left(\nu -\omega_{21}+\Delta\right)-\frac{\Gamma_{31}}{2}-\frac{\gamma_{3}}{2}\right]\left[\mathrm{i}\left(\nu -\omega_{21}\right)
-\frac{\Gamma_{21}}{2}-\frac{\gamma_{2}}{2}\right]+\frac{\Omega^2}{4}},
\\
&&\tilde{\Lambda}_{3}(\nu )=\frac{1}{\sqrt{2\pi}}\frac{-\frac{1}{2}V_{21}\Omega}{\left[\mathrm{i}\left(\nu -\omega_{21}+\Delta\right)-\frac{\Gamma_{31}}{2}-\frac{\gamma_{3}}{2}\right]\left[\mathrm{i}\left(\nu -\omega_{21}\right)
-\frac{\Gamma_{21}}{2}-\frac{\gamma_{2}}{2}\right]+\frac{\Omega^2}{4}}.
\label{freUC}
\end{eqnarray}
\end{widetext}
Here, $\vert \tilde{T}_{b}\vert^2$  gives
the probability of elastic scattering, ( i.e., the outgoing photon experiencing no frequency shift), and  $\vert \tilde{T}_{a}\vert^2$  gives the probability of inelastic scattering (i.e., a photon with up-shifted frequency $\nu+\omega$ is generated). The optimal parameters of the classical driving field for frequency up conversion are 
\begin{equation}
\Delta=0,~~~
\Omega=\sqrt{\left(\Gamma_{31}+\gamma_{3}\right)\left(\Gamma_{21}-\gamma_{2}\right)}.
\label{opt-condition-UC}
\end{equation}
And the resulting transmission coefficients for an on-resonance incoming photon (with $\nu=\omega_{21}$) are
\begin{equation}
|\tilde{T}_{b}(\omega_{21})|^2=0,~~~|\tilde{T}_{a}(\omega_{21})|^2=\frac{1-{\gamma_{2}}/{\Gamma_{21}}}{1+{\gamma_{3}}/{\Gamma_{31}}},
\label{trans-UC}
\end{equation}
meaning that in ideal case $\gamma_{2}=\gamma_{3}=0$, an incoming photon with frequency $\nu$ will be completely converted to a outgoing photon with frequency $\nu+\omega$. 

\subsection{\label{scattering-with-BW}Scattering of a single-photon with finite bandwidth}

\begin{figure*}[t]
\centering
\includegraphics[width=0.9\textwidth]{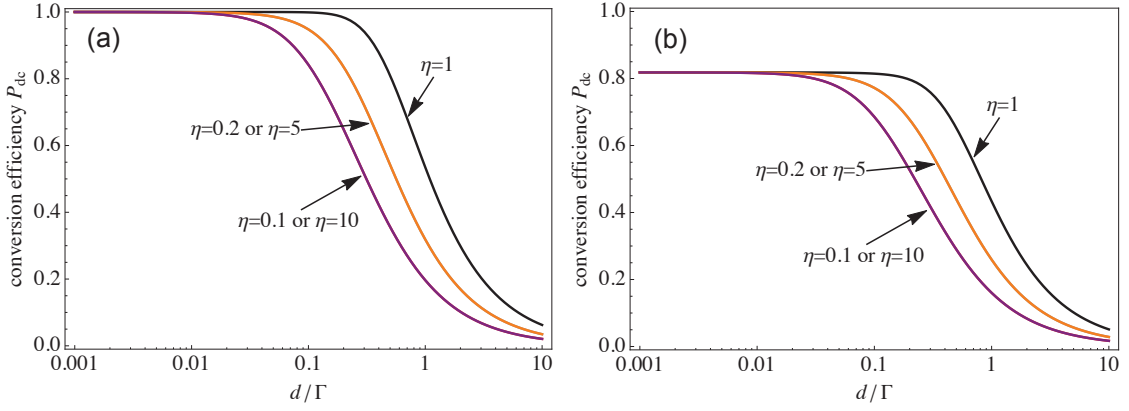}
\caption{(Color online) Frequency down-conversion efficiency for a Gaussian pulse given by Eq.~\eqref{GaussWP} as function of pulse width $d$ for different values of $\eta = \Gamma_{21}/\Gamma_{31}$. (a) The ideal case with no photon loss; (b) The case with photon loss.}
\label{conv_efficient}
\end{figure*}

Now we construct the scattering matrix using the scattering eigenstates given above. Here we take frequency down-conversion case as an example.
According to the Lippmann-Schwinger formalism~\cite{Shen2007PRL, SakuraiBook}, one can show that an input state with single frequency is scattered to an output state 
\begin{equation}
 T_{a}(\nu)\left\vert\nu\right\rangle_{a}
+ T_{b}(\nu)\left\vert\nu-\omega\right\rangle_{b},
\end{equation}
where $T_{a}(\nu)$ and  $T_{b}(\nu)$ are given by Eqs.~\eqref{TranTa} and \eqref{TranTb}, $\left\vert\nu\right\rangle_{a}$ is the input state given by Eq.~\eqref{initial-a}, and $\left\vert\nu-\omega\right\rangle_{b}$ is defined as
\begin{equation}
\left\vert\nu-\omega\right\rangle_{b}=\frac{1}{\sqrt{2\pi}}\int\mathrm{d}x e^{\mathrm{i} (k-\frac{\omega}{v_{\mathrm{g}}}) x-\mathrm{i} (\nu-\omega) t}\hat b_{\mathrm{e}}^{\dagger}\left(x\right)\left\vert\varnothing,1\right\rangle.
\label{final-b} 
\end{equation}
The corresponding scattering matrix can be constructed as .
\begin{equation}
\hat{S}_{\mathrm{dc}}=\int\mathrm{d}\nu\left[T_{a}(\nu)\left\vert\nu\right\rangle_{a}+ T_{b}(\nu)\left\vert\nu-\omega\right\rangle_{b}\right]\left\langle\nu\right\vert_{a}.
\end{equation}

Utilizing scattering matrix given above we can consider a more realistic situation and deal with the problem of scattering of a single-photon pulse with finite bandwidth. A general incoming $a$-mode single-photon state can be written as  
\begin{equation}
\left|\Psi_{\mathrm{in}}\right\rangle=\int\mathrm{d}\nu\psi_{a,\mathrm{in}}(\nu)\left\vert\nu\right\rangle_{a},
\end{equation}
where $\psi_{a,\mathrm{in}}(\nu)$ is the spectral amplitude of the single photon pulse. We assume that the central frequency is near the transition frequency $\omega_{31}$, and the pulse width is much less than the effective atom-waveguide decay rate so that the input photon can be safely looked on as an $a$-mode photon. 
Using the S-matrix defined above,  one can obtain the corresponding output state 
\begin{eqnarray}
\left|\Psi_{\mathrm{out}}\right\rangle&=&\hat{S}_{\mathrm{dc}}\left|\Psi_{\mathrm{in}}\right\rangle
\nonumber
\\
&=&\int\mathrm{d}\nu T_{a}(\nu)\psi_{a,\mathrm{in}}(\nu)\left\vert\nu\right\rangle_{a}
\nonumber
\\
&&+\int\mathrm{d}\nu  T_{b}(\nu)\psi_{a,\mathrm{in}}(\nu)\left\vert\nu-\omega\right\rangle_{b}.
\label{outstatePulse}
\end{eqnarray}
Here, the first term represents the elastic scattered component, and the second term is the inelastic scattered component. Notably, in our proposal, the state of emitter is initially prepared in its ground state $|1\rangle$. After  of single-photon scattering, the emitter again return to its ground state, as shown by the output state~\eqref{outstatePulse}. Thus we can start next frequency-conversion operation without requiring initialization of the atomic state once again.

The efficiency of frequency conversion is defined as the area ratio of the inelastic scattered component to the input pulse. Assuming that the spectral amplitude $\psi_{a,\mathrm{in}}(\nu)$ has been normalized, i.e., $\int\mathrm{d}\nu |\psi_{a,\mathrm{in}}(\nu)|^2=1$, the efficiency of frequency conversion takes the form
\begin{equation}
P_{\mathrm{dc}}=\int\mathrm{d}\nu  \left| T_{b}(\nu)\psi_{a,\mathrm{in}}(\nu)\right |^2.
\label{Pdc}
\end{equation}

As shown in Sec.~\ref{photon-scattering}, the contribution from off-resonance frequencies degrades conversion efficiency, thus to obtain a conversion efficiency close to $1$, first, the central frequency of the wave packet should be on-resonance with the transition channel $|3\rangle\leftrightarrow|1\rangle$, and second, the pulse width $d$ of $|\psi_{a,\mathrm{in}}|^2$ should be much less than the width of transmission spectra $|T_{a}|^2$ and $|T_{b}|^2$. Based on the results in Sec.~\ref{photon-scattering}, this can be summarized as
\begin{equation}
d\ll\min(\Gamma_{31},\Gamma_{21}).
\label{HEcondition}
\end{equation}
Under these conditions, $|\psi_{a,\mathrm{in}}|^2$ can be approximated as a $\delta$-function, namely, $|\psi_{a,\mathrm{in}}|^2\sim\delta(\nu-\omega_{31})$, thus according to Eq.~\eqref{Pdc}, $P_{\mathrm{dc}}\simeq |T_{b}(\omega_{31})|^2$. In this case of quasi-monochromatic input, if the losses through dissipation are negligible, then we have $|T_{b}(\omega_{31})|^2=1$ [see Fig.~\ref{TransmissionSpectrum}(a)-(c)]. This means that conversion efficiency close to unity is possible.
Note that condition~\eqref{HEcondition} can also guarantee that the pulse widths of the incoming and outgoing photons are much less than the effective atom-waveguide decay rate, so that they can be safely treated as $a$-mode and $b$-mode photons, respectively. 

As an example, we take a wave packet with Gaussian-type spectral amplitude
\begin{equation}
\psi_{a,\mathrm{in}}(\nu)={\left(\frac{2}{\pi d^2}\right)}^{\frac{1}{4}}e^{-\frac{\left(\nu-\omega_{31}\right)^2}{d^2}},
\label{GaussWP}
\end{equation}
where $d$ is the pulse width. Fig.~\ref{conv_efficient}(a) and~\ref{conv_efficient}(b) shows the efficiency of frequency conversion for a
Gaussian pulse given by Eq.~\eqref{GaussWP} as function of pulse width $d$ for fixed $\Gamma=\Gamma_{21}+\Gamma_{31}$ and different
$\eta=\Gamma_{21}/\Gamma_{31}$. Clearly, when  the photon loss is not included and the pulse width $d$
satisfies the condition~\eqref{HEcondition}, a conversion efficiency close to unity can be achieved
[Fig.~\ref{conv_efficient}(a)], which is in accordance with our earlier analysis. While Fig.~\ref{conv_efficient}(b) 
shows the efficiency of frequency conversion when low photon loss is included. In this case, for a sufficiently 
narrow photon pulse satisfying condition~\eqref{HEcondition}, according to Eq.~\eqref{trans-DC}, the efficiency can be
 approximated as $P_{\mathrm{dc}}\simeq|T_{b}(\omega_{31})|^2\simeq 1-\gamma_{2}/\Gamma_{21}-\gamma_{3}/\Gamma_{31}$, which is verified by Fig.~\ref{conv_efficient}(b).

At the end of this section, we give the main results of frequency up-conversion case. The scattering matrix in this case is
\begin{equation}
\hat{S}_{\mathrm{uc}}=\int\mathrm{d}\nu \left[ \tilde{T}_{a}(\nu )\left\vert\nu  +\omega\right\rangle_{a}+ \tilde{T}_{b}(\nu )\left\vert\nu \right\rangle_{b}\right]\left\langle\nu \right\vert_{b}.
\end{equation}
For an incoming $b$-mode single-photon state
\begin{equation}
\left|\Psi_{\mathrm{in}}\right\rangle=\int\mathrm{d}\nu\psi_{b,\mathrm{in}}(\nu)\left\vert\nu\right\rangle_{b},
\end{equation}
with normalized spectral amplitude $\psi_{b,\mathrm{in}}(\nu)$. The corresponding efficiency of frequency conversion is the area ratio of the inelastic scattered component to the input pulse and has the form
\begin{equation}
P_{\mathrm{uc}}=\int\mathrm{d}\nu \left| \tilde{T}_{a}(\nu)\psi_{b,\mathrm{in}}(\nu)\right |^2.
\end{equation}

\section{\label{Experimental-realization}Experimental realization in superconducting quantum circuits}

\subsection{\label{Eff-Hamiltonian}Effective Hamiltonian for superconducting quantum circuit }

\begin{figure}[h]
\centering
\includegraphics[width=0.5\textwidth]{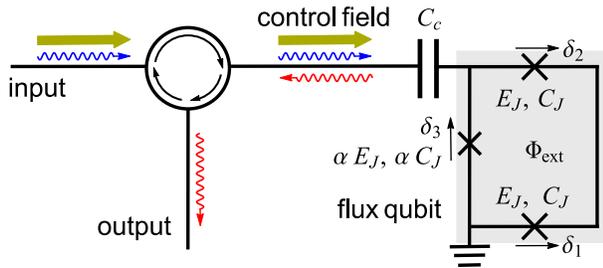}
\caption{(Color online) 
The circuit diagram of a three-junction flux qubit capacitively coupled to the end of a semi-infinite transmission line. The frequency-shifted photon can be measured in reflection. A circulator is used to separate the input and output fields.}
\label{circuit}
\end{figure}

We now further study our proposal by considering a more concrete example. We specify our three level superconducting quantum system to a  flux qubit (also called artificial atom), which is embedded at the end of a semi-infinite one-dimensional transmission line, as sketched in Fig.~\ref{circuit}. The flux qubit circuit consists of a superconducting loop with three Josephson junctions. The two larger ones are identical and have the same Josephson energies $E_{\mathrm{J1}}=E_{\mathrm{J2}}=E_{\mathrm{J}}$ and capacitances $C_{\mathrm{J1}}=C_{\mathrm{J2}}=C_{\mathrm{J}}$, while for the third junction  $E_{\mathrm{J3}}=\alpha E_{\mathrm{J}}$ and $C_{\mathrm{J3}}=\alpha C_{\mathrm{J}}$, with 􏻀 $\alpha<1$. The semi-infinite transmission line has characteristic inductance $l$ and capacitance $c$ per unit length. The qubit circuit is coupled to the  semi-infinite transmission line through a coupling capacitance $C_{\mathrm{c}}$. The incident single photon and the control microwave field are applied through the input port. A circulator is used to separate the input and output photons. 

Following the method used in Refs.~\cite{Delsing2015NATPHYS,CapacityCoupl1,CapacityCoupl2}, we derive the Hamiltonian of the full system, which consists of the flux qubit, the transmission line and the coupling parts. The Hamiltonian of flux qubit part is given by
\begin{eqnarray}
\hat{H}_{\mathrm{atom}}&=&E_{\mathrm{C}}\hat{n}_{\mathrm{p}}^{2}+\frac{2E_{\mathrm{C}}}{1+2\alpha+2\beta}\hat{n}_{\mathrm{m}}^{2}
\nonumber
\\
&&+E_{\mathrm{J}}\left[2+\alpha-2\cos\delta_{\mathrm{p}}\cos\delta_{\mathrm{m}}\right.
\nonumber
\\
&&\left.-2\alpha\cos\left(2\delta_{\mathrm{m}}+2\pi f\right)\right],
\label{Fluxqubit}
\end{eqnarray}
where  $\beta=C_{\mathrm{c}}/C_{\mathrm{J}}$, $E_{\mathrm{C}}=e^2/(2C_{\mathrm{J}})$ is the charging energy.
$\delta_{\mathrm{p}}=(\delta_{\mathrm{1}}+\delta_{\mathrm{2}})/2$ and $\delta_{\mathrm{m}}=(\delta_{\mathrm{1}}-\delta_{\mathrm{2}})/2$ are defined by the phase drops $\delta_{\mathrm{1}}$ and $\delta_{\mathrm{2}}$ across the two larger junctions. The  charge number operators $\hat{n}_{\mathrm{p}}=-\mathrm{i}\frac{\partial}{\partial\delta_{\mathrm{p}}}$ and $\hat{n}_{\mathrm{m}}=-\mathrm{i}\frac{\partial}{\partial\delta_{\mathrm{m}}}$
are conjugate variables of $\delta_{\mathrm{p}}$ and $\delta_{\mathrm{m}}$.  $f=\Phi_{\mathrm{ext}}/\Phi_{0}$ is the reduced magnetic flux. Here $\Phi_{\mathrm{ext}}$˜ is an external magnetic flux through the qubit loop and $\Phi_{0}=h/(2e)$ is the flux quantum. We choose the lowest three energy levels $|1\rangle$, $|2\rangle$ and $|3\rangle$, which can form a $\Delta$-type three-level artificial atom when $f{\neq}0.5$~\cite{DeltaAA1}, to implement our frequency conversion scheme. Using the three lowest eigen states, one can truncate the Hilbert space of the atom part into three dimensions and rewrite the atomic Hamiltonian \eqref{Fluxqubit} as
\begin{equation}
\hat{H}_{\mathrm{atom}}=\sum_{i=2,3}\hbar\omega_{i1}\left\vert i\right\rangle \left\langle i\right\vert.
\end{equation}  
Here the energy of the ground state $\left\vert 1\right\rangle$ is set to zero as reference.

Let us now consider the free Hamiltonian of the transmission line.  We assume that the transitions frequencies greatly exceed the line widths, which can be verified by the following numerical simulations using experimentally feasible parameters in Sec.~\ref{Numerical}. We also assumed that the difference between $ \omega_{32}$ and $\omega_{21}$ are exceed the line widths. Thus we can treat the transmission-line modes as three different quasi-monochromatic modes. The photon modes coupling the transitions $|1\rangle\leftrightarrow |3\rangle$ and $|1\rangle\leftrightarrow |2\rangle$ are relevant to input and output single photons, thus we treat them as quantized fields. While the strong driving filed coupling the energy levels $|2\rangle$ and $|3\rangle$ is treated as classical one. The two quantized photon modes is described by the Hamiltonian 
\begin{equation}
\hat{H}_{\mathrm{WG}}=\int_{k\simeq k_{31}}\mathrm{d}k\hbar\omega_{k}\hat{a}_{k}^{\dagger}\hat{a}_{k}+\int_{k\simeq k_{21}}\mathrm{d}k\hbar\omega_{k}\hat{b}_{k}^{\dagger}\hat{b}_{k},
\label{TL}
\end{equation}
with bosonic creation (annihilation) operators $\hat{a}_{k}^{\dagger}$($\hat{a}_{k}$) and $\hat{b}_{k}^{\dagger}$($\hat{b}_{k}$). Here we have mapped the left- and right-modes in a semi-infinite transmission line to a unidirectional transporting chiral modes in a infinite transmission line, as shown in Fig.\ref{Fig.sub.1}. We are only interested in photons with narrow bandwidths in the vicinity of atomic transition frequencies, thus the integrals in \eqref{TL} should be carried out over narrow intervals around $k_{31}$ and $k_{21}$, with corresponding frequency $\omega_{31}$ and $\omega_{21}$, respectively.

Under rotating wave approximation, the interaction between the propagating photons and the flux qubit is governed by
\begin{eqnarray}
\hat{H}_{\mathrm{int}}&=&-\frac{\hbar V_{31}}{\sqrt{2\pi}}\int_{k\simeq k_{31}}\mathrm{d}k\left(\hat{a}_{k}^{\dagger}\left |1\right\rangle\left\langle 3\right |+\mathrm{H.c.}\right)
\nonumber
\\
&&-\frac{\hbar V_{21}}{\sqrt{2\pi}}\int_{k\simeq k_{21}}\mathrm{d}k\left(\hat{b}_{k}^{\dagger}\left |1\right\rangle\left\langle 2\right |+\mathrm{H.c.}\right),
\label{fluxTL}
\end{eqnarray}
where
\begin{equation}
V_{ij}=\frac{1}{\hbar}\frac{2e \beta}{1+2 \alpha+2 \beta}\sqrt{\frac{2\hbar\omega_{ij}}{c}}{\left\vert 
n_{ij}\right\vert}
\end{equation}
is the coupling strength between the flux qubit and quantized $a$- and $b$-mode photons in the transmission line. $n_{ij}=\left\langle i\right|\hat{n}_{\mathrm{m}}\left| j\right\rangle$ is the transition elements.  In general, $V_{ij}$ is $\omega_{k}$ dependent. However, in our case only photons within a narrow bandwidth around the atomic frequency $\omega_{ij}$ can effectively couple to the qubit, therefore we approximate the coupling constant $V_{ij}$ as its value at $\omega_{k}=\omega_{ij}$. This assumption is equivalent to the Markovian approximation~\cite{Gardiner1985PRA}. Using the Fermi Golden Rule~\cite{Delsing2015NATPHYS, Clerk2010RMP}, one can verify that the interaction Hamiltonian \eqref{fluxTL} gives rise to the rate of spontaneous emission into the transmission line as follows
\begin{equation}
\Gamma_{ij}=\frac{2} {\hbar} \left(\frac{2e \beta}{1+2 \alpha+2 \beta}\right)^{2}Z \omega_{ij} {\left\vert
n_{ij}\right\vert}^{2},
\end{equation}
where $Z=\sqrt{l/c}$  is the characteristic impedance of the transmission line. 

Finally, we give the Hamiltonian describing interaction between the classical field and the flux qubit. In the presence of a classical microwave field, the voltage felt by the qubit locating at the end of the transmission line is $\frac{1}{2}\mathcal{V}_\mathrm{c}(\mathrm{e}^{-i\omega t}+\mathrm{c.c.})$, where $\mathcal{V}_\mathrm{c}$ and $\omega$ are the amplitude and frequency of the classical field, respectively. Here, we assume $\mathcal{V}_\mathrm{c}$ is real and $\omega\simeq\omega_{32}$.   Under rotating wave approximation, the corresponding interaction Hamiltonian can be written as
\begin{equation}
\hat{H}_{\mathrm{d}}=-\frac{\hbar\Omega}{2}\left(\left\vert 3\right\rangle \left\langle 2\right\vert \mathrm{e}^{-\mathrm{i}\omega t}+\mathrm{H.c.}\right),
\label{ClassicDriving}
\end{equation}
where
\begin{equation}
\Omega=\frac{1}{\hbar}\frac{2e \beta}{1+2 \alpha+2 \beta}\left\vert n_{32}\right\vert\mathcal{V}_\mathrm{c}
\end{equation}
is the Rabi frequency of the classical field.

We assume that in the entire frequency range of interest the linear dispersion relation holds. Thus the frequency of $a$- and $b$-mode photon can be written as $\omega_{k}=\omega_{a}+v_{\mathrm{g}}(k-k_{a})$ and $\omega_{k}=\omega_{b}+v_{\mathrm{g}}(k-k_{b})$, respectively, where $\omega_{a}$ ($\omega_{b}$) is some arbitrary frequency near $\omega_{31}$ ($\omega_{21}$), $k_{a}$ ($k_{b}$) is the corresponding wave vector. $v_{\mathrm{g}}$ is the group velocity of the propagating photons.
In a rotating reference frame defined by the unitary transformation, 
\begin{eqnarray}
U&=&\exp\left[\mathrm{i}\omega_{a}\left(\int_{k\simeq k_{31}}\mathrm{d}k\hat{a}_{k}^{\dagger}\hat{a}_{k}+|3\rangle\langle 3|\right)t\right.
\nonumber
\\
&&\left.+\mathrm{i}\omega_{b}\left(\int_{k\simeq k_{21}}\mathrm{d}k\hat{b}_{k}^{\dagger}\hat{b}_{k}+|2\rangle\langle 2|\right)t\right],
\end{eqnarray}
 the total Hamiltonian can be written as
\begin{eqnarray}
\hat{H}/\hbar&=&\int_{-\infty}^{+\infty}\mathrm{d}k' v_{\mathrm{g}}k'\hat{a}_{k'}^{\dagger}\hat{a}_{k'}+\int_{-\infty}^{+\infty}\mathrm{d}k'' v_{\mathrm{g}}k''\hat{b}_{k''}^{\dagger}\hat{b}_{k''}
\nonumber
\\
&&+\left(\omega_{31}-\omega_{a}\right)\left\vert 3\right\rangle \left\langle 3\right\vert+\left(\omega_{21}-\omega_{b}\right)\left\vert 2\right\rangle \left\langle 2\right\vert
\nonumber
\\
&&-\frac{\Omega}{2}\left(\left\vert 3\right\rangle \left\langle 2\right\vert \mathrm{e}^{-\mathrm{i}[\omega-(\omega_{a}-\omega_{b})] t}+\mathrm{H.c.}\right)
\nonumber
\\
&&- \frac{V_{31}}{\sqrt{2\pi}}\int_{-\infty}^{+\infty}\mathrm{d}k'\left(\hat{a}_{k'}^{\dagger}\left |1\right\rangle\left\langle 3\right |+\mathrm{H.c.}\right)
\nonumber
\\
&&-\frac{V_{21}}{\sqrt{2\pi}}\int_{-\infty}^{+\infty}\mathrm{d}k''\left(\hat{b}_{k''}^{\dagger}\left |1\right\rangle\left\langle 2\right |+\mathrm{H.c.}\right).
\label{QubitTLHamiltonian}
\end{eqnarray}
Here we have set $k'=k-k_{a},~k''=k-k_{b}$ and extended the limits of the sum over $k'$ ($k''$)  to $(-\infty, \infty)$.
By defining
\begin{eqnarray}
\hat{a}_{k'}=\int_{-\infty}^{+\infty}\mathrm{d}x \hat{a}_{\mathrm{e}}(x){\mathrm{e}}^{-\mathrm{i}k'x},
\\
\hat{b}_{k''}=\int_{-\infty}^{+\infty}\mathrm{d}x \hat{b}_{\mathrm{e}}(x){\mathrm{e}}^{-\mathrm{i}k''x},
\end{eqnarray}
and including the dissipation rate $\gamma_{i}$ of the atom (including the contributions of pure dephasing and photon loss to other modes), one can get
the effective real-space Hamiltonian \eqref{Hamiltonian}. Note that in Hamiltonian \eqref{Hamiltonian}, we have adopted renormalized frequencies by making a change of variables $\omega_{31}-\omega_{a}\rightarrow\omega_{31}$, $\omega_{21}-\omega_{b}\rightarrow\omega_{21}$, and $\omega-(\omega_{a}-\omega_{b})\rightarrow\omega$. 
\begin{figure*}[t]
\centering
\includegraphics[width=0.9\textwidth]{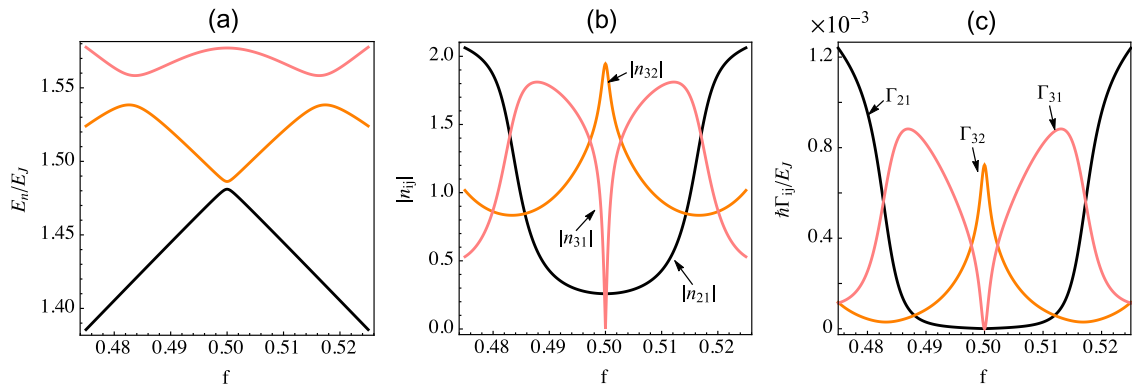}
\caption{(Color online) Flux bias dependence of energy levels,  transition matrix elements, and spontaneous emission rates, for the lowest three states. The circuit parameters are chosen as $\alpha=0.7$, $\beta=0.5$, $E_{\mathrm{J}} =80E_{\mathrm{C}}$ and $Z=50~\Omega$. (a) Energy levels, in units of $E_{\mathrm{J}}$, of the flux qubit vs reduced flux $f$. (b) Moduli $|n_{ij}|$ of the transition matrix elements between states $|i\rangle$ and $|j\rangle$ vs $f$. (c) Spontaneous emission rates, in units of $E_{\mathrm{J}}$, between states $|i\rangle$ and $|j\rangle$ vs $f$.}
\label{ELandTE}
\end{figure*}

\subsection{\label{Numerical}Numerical results with experimentally feasible parameters}

Figure.~\ref{ELandTE}(a) shows the energy spectrum for the qubit Hamiltonian~\eqref{Fluxqubit} and 
Fig.~\ref{ELandTE}(b) the corresponding transition matrix elements, both as functions of reduced magnetic flux 
$f$. The circuit parameters are chosen as $\alpha=0.7$, $\beta=0.5$, $E_{\mathrm{J}}/\hbar=2\pi\times150~\mathrm{GHz}$, $E_{\mathrm{C}} =E_{\mathrm{J}}/80$ and $Z=50~\Omega$, which are experimentally feasible~\cite{fluxqubit,CapacityCoupl1,CapacityCoupl2}.
One can find that the transition matrix elements have comparable values when the reduced flux is slightly 
away from the degenerate point $f=0.5$, all the three transitions are nonzero, the flux qubit can then be used as a 
$\Delta$-type artificial atom~\cite{DeltaAA1} to realize the scheme shown in Fig~\ref{WG_ATOM}(b). Note that the spacing between energy levels is highly tunable by adjusting the flux bias, enabling a broadband dynamic range in our frequency convertor.

Figure.~\ref{ELandTE}(c) shows the spontaneous emission rate $\Gamma_{ij}$ as a function of reduced magnetic 
flux $f$. One can see that $\Gamma_{ij}$ are about three orders of magnitude smaller than the energy 
separations of the three-level system. Thus our previous  treating the transmission line modes as distinct ones is reasonable. In 
Sec~\ref{photon-scattering}. we have shown that for both frequency up- and down-conversion cases, to obtain an 
efficiency close to 1, sufficiently low photon loss should be guaranteed, namely, the condition~\eqref{DecayCondition} 
should be satisfied. Note that in condition~\eqref{DecayCondition}, $\gamma_{3}$ models the spontaneous emission of the atomic level
$|3\rangle$ due to coupling to the degrees of freedom other than $a$-modes. Specifically, these including the degrees of freedom in the environment and the transmission-line modes coupling to the transition $|2\rangle\leftrightarrow |3\rangle$, where the corresponding decay rates are defined as $\tilde{\gamma}_{3}$ and $\Gamma_{32}$, respectively, satisfying $\gamma_{3}=\tilde{\gamma}_{3}+\Gamma_{32}$. While $\gamma_{2}$ describes the
spontaneous emission of the atomic level $|2\rangle$ due to coupling to modes other than $b$-modes. To satisfy 
condition \eqref{DecayCondition}, we suggest that the emitter is perfectly coupled to the transmission line, i.e., 
$\gamma_{2}\simeq 0$, $\tilde{\gamma}_{3}\simeq 0$, and at the same time $\Gamma_{32}\ll \Gamma_{31}$. 
These conditions are experimentally feasible in our proposal based on superconducting circuits. Firstly, recent 
experiments~\cite{Tsai2010Science,Delsing2011PRL,Delsing2012PRL,Delsing2013PRL} have demonstrated that the microwave photons can be coupled extremely efficiently to a single 
artificial atom, showing extinction efficiencies in an open transmission line up to $99.6\%$~\cite{Delsing2011PRL}. 
That is to say, the majority of the decayed light from the artificial atom is guided into transmission line modes (so called ``strong coupling'' between the atom and the transmission line~\cite{Shen2005PRL,Chang2007NatPhys}), meaning that $\gamma_{2}\simeq 0$, 
$\tilde{\gamma}_{3}\simeq 0$.  In addition, the condition $\Gamma_{32}\ll \Gamma_{31}$ can be guaranteed by
biasing the flux qubit around $f=0.485$, or $f=0.515$, as shown in Fig.~\ref{ELandTE}(c).

\begin{figure*}[t]
\centering
\includegraphics[width=0.75\textwidth]{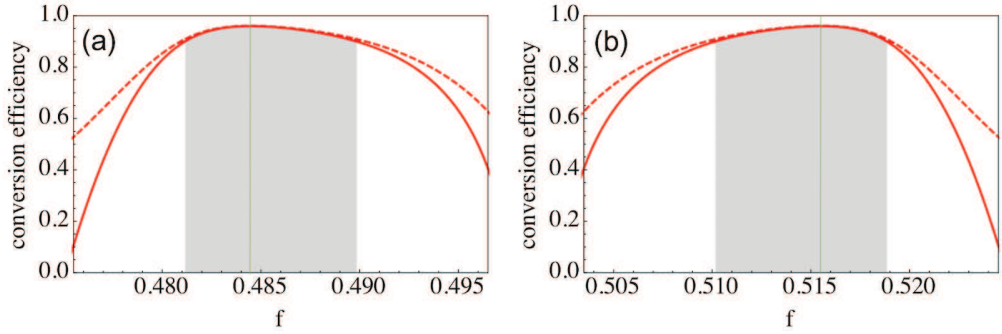}
\caption{(Color online) The conversion efficiencies for a monochromatic and on-resonance photon as functions of $f$. Here perfect coupling between flux qubit and transmission line is assumed. The solid line is the down-conversion efficiency, and the dashed line plots the up-conversion efficiency. The flux bias $f = 0.4845$ and $f = 0.5155$, marked by the vertical thin lines in (a) and (b), are optimal working points for frequency conversion. (a) The conversion efficiency around $f = 0.4845$; (b) The conversion efficiency around $f = 0.5155$.}
\label{simexp1}
\end{figure*}
\begin{figure}[b]
\centering
\includegraphics[width=0.5\textwidth]{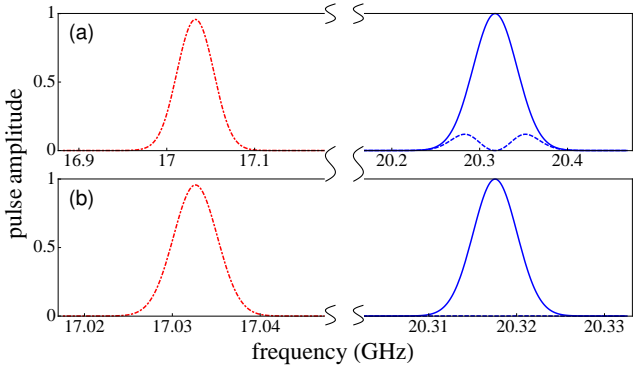}
\caption{(Color online). Input and output pulses for the case of frequency down conversion. The input Gaussian pulses with different pulse widths, shown by the solid blue curves, are centered at frequency $20.318~\mathrm{GHz}$ : (a) The width of input pulse is $0.05~\mathrm{GHz}$; (b) The width of input pulse is $0.005~\mathrm{GHz}$. Dashed blue curves are the elastically scattered (i.e., frequency-unshifted) outputs, and dot-dashed red curves are the inelastically scattered
(i.e., frequency-shifted) outputs. The flux bias is set as $f = 0.4845$, the qubit-waveguide coupling rate is $99.9\%$, other circuit parameters are the same as in Fig.~\eqref{ELandTE}. }
\label{pulse}
\end{figure}

To verify above analysis, we plot the conversion efficiencies for perfect coupling case (i.e., $\gamma_{2}\simeq 0, \gamma_{3}\simeq\Gamma_{32}$) as functions of $f$ around these two areas, as shown in Figs.~\ref{simexp1}(a) and \ref{simexp1}(b). The solid line is the conversion efficiency for frequency down-conversion case, and the
dashed line for frequency up-conversion case, respectively. Here we assume the photon is monochromatic and on-resonance, the according conversion efficiencies are given by $|T_{b}|^2$ in Eq.~\eqref{trans-DC} (frequency down-conversion case) and $|\tilde{T}_{a}|^2$ in Eq.~\eqref{trans-UC} (frequency up-conversion case), respectively. Clearly, these results give the upper limits (i.e., single frequency limits) for the case of a single-photon pulse with finite bandwidth. Specifically, the flux bias $f=0.4845$ and $f=0.5155$, indicated by the vertical thin lines in Figs.~\ref{simexp1}(a) and \ref{simexp1}(b), are optimal for frequency conversion. The corresponding conversion efficiencies are $95.9\%$ for frequency down-conversion case (solid line), and $96.1\%$ for frequency up-conversion case (dashed line), respectively.  In addition, for both down-conversion and up-conversion cases, efficiencies over $90\%$ can be obtained in a desirable flux bias range $0.4812 < f < 0.4898$ or $0.5102 < f < 0.5188$, as shown by the shaded areas in Figs.~\ref{simexp1}(a) and \ref{simexp1}(b).

When biasing the flux in this region, while $\Gamma_{32}$ is much less than $\Gamma_{31}$, but the corresponding transition elements $|n_{32}|$ is not so small, as shown in Fig.~\ref{ELandTE}(b), thus energy levels $|2\rangle$ and $|3\rangle$ can be easily coupled by applying a classical driving field. 
Our simulation shows that, if the circuit parameters are the same as those in Fig.~\ref{ELandTE},  to obtain a Rabi frequency satisfying the condition \eqref{opt-condition-DC} or \eqref{opt-condition-UC}, the order of magnitude of the amplitude of  the classical field should be $10^{-7}\sim 10^{-6}~\mathrm{V}$, which is experimentally feasible~\cite{Wallraff2004Nature}. 

In practice, a single-photon pulse is always with finite bandwidth. Fig.~\ref{pulse} shows input Gaussian photon pulses and resulted outputs, illustrated by the case of frequency down-conversion. The flux bias is set as $f=0.4845$, which is optimal for frequency conversion. The corresponding frequency separations of qubit are $\omega_{31}/2\pi=20.318~\mathrm{GHz}$, $\omega_{21}/2\pi=17.033~\mathrm{GHz}$ and $\omega_{32}/2\pi=3.285~\mathrm{GHz}$, and the spontaneous decay rates are $\Gamma_{31}/2\pi=0.118~\mathrm{GHz}$, $\Gamma_{21}/2\pi=0.041~\mathrm{GHz}$, and $\Gamma_{32}/2\pi=0.005~\mathrm{GHz}$. According to recent experiments~\cite{Delsing2011PRL}, we assume that $\tilde{\gamma}_{3}/\Gamma_{31}=\gamma_{2}/\Gamma_{21}=0.001$. That is to say, the qubit-waveguide coupling efficiency is about $99.9\%$. In Figs.~\ref{pulse}(a) and \ref{pulse}(b), the solid blue curves show the input pulses centered at frequency $20.318~\mathrm{GHz}$, which is on resonance with the $|1\rangle\leftrightarrow|3\rangle$ transition. Dashed blue curves are the elastically scattered (i.e., frequency-unshifted) outputs, and dot-dashed red curves are the inelastically scattered (i.e., frequency-shifted) outputs. In Fig.~\ref{pulse}(a) the width of input pulse is $0.05~\mathrm{GHz}$, which is comparable to $\Gamma_{21}$. In this case, the condition \eqref{HEcondition} is not satisfied. The frequency-unshifted outputs are not negligible, as shown by the dashed blue curve in Fig~\ref{pulse}(a). Consequently, the conversion efficiency is lowered to $78.6\%$ due to off-resonance effects, far below the single frequency limit $95.9\%$. On the contrary, in Fig.~\ref{pulse}(b) the width of input pulse is $0.005~\mathrm{GHz}$, satisfying condition \eqref{HEcondition}.  In this case,  the elastically scattered outputs almost vanish [see the dashed blue curve in Fig.~\ref{pulse}(b)] because the off-resonance effects can be negligible. Accordingly, the conversion efficiency is up to $95.5\%$, only slightly below the single frequency limit $95.9\%$. 

\begin{figure*}[t]
\centering
\includegraphics[width=\textwidth]{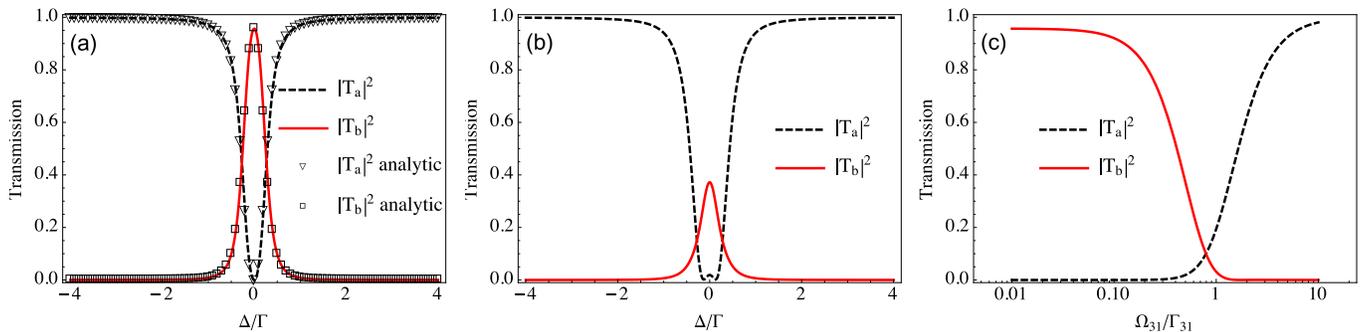}
\caption{(Color online) (Color online) Effects of nonlinearity on  the  transmission spectra (frequency down-conversion case). (a) Transmission spectra when $\Omega_\mathrm{p}=0.01\Gamma_{31}$; (b) Transmission spectra when $\Omega_\mathrm{p}=0.5\Gamma_{31}$; (c) Transmission coefficients as function of $\Omega_\mathrm{p}$. Other parameters are the same as in Fig.~\eqref{pulse}.}
\label{nonlinear}
\end{figure*}

In experiments, the input field can also be a weak continuous microwave field (probe field), instead of a single-photon. If the input continuous field is sufficiently weak (single-photon limit), our analytical calculations based on single-photon scattering problem can still well describe the frequency-conversion process. If the input power is increased, the conversion efficiency becomes lower due to saturation of the atom. In this nonlinear region, we can treat the problem numerically. Here we take the case of frequency down-conversion as an example.  We can write down a semi-classical Hamiltonian by assuming the input (coupling levels $|1\rangle$ and $|3\rangle$) and control fields (coupling levels $|2\rangle$ and $|3\rangle$) as two continuous microwave fields with Rabi frequencies $\Omega_\mathrm{p}$ and $\Omega$, respectively. We can obtain the transmission coefficients (frequency-conversion efficiency) of the input field using master equation method (see Appendix for details). In our simulation, the circuit parameters are chosen as those in Fig~\ref{pulse}.  In Fig.~\ref{nonlinear}(a), we compare the analytical results [Eq.~\eqref {TranTa} and Eq.~\eqref {TranTb}] with numerical ones under condition $\Omega_\mathrm{p}\ll\Gamma_{31}$. We can see that when the input continuous microwave field is sufficiently weak ($\Omega_\mathrm{p}=0.01\Gamma_{31}$), the numerical results are in good agreement with the analytical ones obtained by treating the input field quantum mechanically. Clearly, in this weak field (or single-photon) limit, a high conversion efficiency close to 1 can be obtained, as shown in Fig.~\ref{nonlinear}(a). When the power of the incident microwave is increased ($\Omega_\mathrm{p}\sim\Gamma_{31}$ or $\Omega_\mathrm{p}>\Gamma_{31}$), the atom will be saturated, resulting a lower transmission probability $|T_b|^2$ (i.e., down-conversion efficiency). Note that in this nonlinear region, the analytical calculations by taking single-photon limit fail. But the transmission spectra can still be numerically calculated using master equation method. Fig.~\ref{nonlinear}(b) shows the transmission spectra when  $\Omega_\mathrm{p}=0.5\Gamma_{31} $. One can see that for a resonantly incident field, the conversion efficiency $|T_b|^2\simeq{37.2\%}$.  The transmission coefficients  for  on-resonance input fields as functions of $\Omega_\mathrm{p}$ are given in Fig.~\ref{nonlinear}(c). With increasing $\Omega_\mathrm{p}$, more photons of the microwave field $\Omega_\mathrm{p}$  transmit without interaction with the atom because of the saturation of the atom excitation. Consequently, the transmission  $|T_a|^2$ monotonically increases, as shown by the dashed line in Fig.~\ref{nonlinear}(c). On the other hand, only the photons interacting with the atom can convert to frequency-shifted ones, thus with increasing 
$\Omega_\mathrm{p}$ the ratio of down-converted photons to input photons must become smaller. Consequently, the transmission probability $|T_b|^2$ monotonically decreases, as shown by the solid line in Fig.~\ref{nonlinear}(c). We should emphasize that, when single-photon condition $\Omega_\mathrm{p}\ll\Gamma_{31}$ is satisfied, the system we studied can work as an idea frequency convertor with conversion efficiency close to 1.  

\section{\label{Conclusions}Conclusions and discussions}

In summary, we study an efficient single-photon frequency conversion in microwave domain based on superconducting quantum circuits. Our proposal requires a single three-level superconducting artificial atom with $\Delta$-type transition configuration embedded at the end of a semi-infinite one-dimensional transmission line. The frequency conversion can be controlled and optimized by tuning the parameters (strength and detuning) of the applied microwave field. We demonstrate that this device can achieve single-photon frequency up- or down-conversion with efficiency close to $100\%$. 

As an example, we study the frequency conversion using superconducting flux qubit circuits. We show that conversion efficiency higher than $95\%$ can be obtained with experimentally feasible parameters. Note that throughout our proposal, the state of emitter only needs to be manually prepared to ground state once. After frequency-conversion operation, the emitter will return to its ground state and is ready for the next operation cycle. Our convertor also works in a broadband frequency range. For example, by changing the flux bias of a flux qubit circuit, the frequencies of input and output photons, and accordingly, the difference between them are highly tunable (a few GHz). The device studied here is suitable for on-chip integrations and may have broad applications in areas such as quantum information processing utilizing superconducting circuits, microwave single-photon detection, and quantum interface connecting devices operating at different frequencies in the future hybrid quantum network.

In our proposal, a control field $\Omega$ is used to optimize the conversion efficiency. We note that the quantum fluctuation of the control field should in principle be included when the strength of the control field is finite. However, in the parameter regime that we discussed, we find that the finite amplitude effect is negligibly small when the Rabi frequency of control field is comparable to the decay rates. Similar to the derivation for input field in Appendix,  we can derive that the average photon-number of the control field within decay time scale $2\pi/\Gamma_{32}$ is $N=\pi\Omega^2/(2\Gamma_{32}^2)$. If we let the Rabi frequency of control field satisfy Eq.~\eqref{opt-condition-DC} and use experimentally feasible parameter given in Sec.~\ref{Numerical}, we find $N\simeq37$, which is much larger than 1. Note that only when $N\ll1$ the control field should be treated as single-photon field and the quantum fluctuation of the control field play significant role. Thus we can safely neglect the fluctuation of the control field within the parameter regime that we considered.

We emphasize that although our discussions on physical realization of our proposal are focused on superconducting circuits, the general analysis in Sec.~\ref{model} can also be applicable for other quantum systems with $\Delta$-type transition structure,  including chiral molecules~\cite{DeltaMolecule1, DeltaMolecule2, DeltaMolecule3}, asymmetric quantum wells~\cite{DeltaMolecule1}, and natural atoms with the two metastable states coupling by a microwave through magnetic-dipole transition~\cite{AtomloopScully}. Thus, the frequency convertor based on our scheme can be implemented in a variety of systems and work at other frequencies besides microwave. 

\begin{acknowledgments}
W. Z. J. is supported by the NSFC under Grant No.~11404269, 11347001, 11647314 and 11547311. Y. W. W. is supported by the NSFC under Grant No.~61301031. Y. X. L. is supported by the NSFC under Grant No.~91321208 and National Basic Research Program under Grand No,~2014CB921401.  
\end{acknowledgments}

\appendix

\section{NUMERICAL CALCULATION USING MASTER EQUATION}

In this appendix, we numerically calculate the frequency conversion efficiency using a semiclassical method. Here we discuss the frequency down-conversion case only.
We treat both the  weak input field (couples levels $|1\rangle$ and $|3\rangle$) and the strong control field (couples levels $|2\rangle$ and $|3\rangle$) as continuous microwave fields.  Note that the end of the transmission line is broken at $x = 0$. Under this boundary condition, the external voltage signal (not including the fields re-emitted by the atom) in the semi-infinite transmission line can be written as 
\begin{eqnarray}
&&V_0(x,t)=\frac{1}{4}\mathcal{V}_{\mathrm{p}}(e^{\mathrm{i}kx}+e^{-\mathrm{i}kx})e^{-\mathrm{i}\nu t}\theta(-x)
\nonumber
\\
&&+\frac{1}{4}\mathcal{V}_\mathrm{c}(e^{\mathrm{i}k_{\omega}x}+e^{-\mathrm{i}k_{\omega} x})e^{-\mathrm{i}\omega t}\theta(-x)+\mathrm{c}.\mathrm{c}.,
\end{eqnarray}
which includes incident field and directly reflected field. Here, without loss of generality, we assume that the amplitudes $\mathcal{V}_{\mathrm{p}}$ and $\mathcal{V}_{\mathrm{c}}$ are real.  The voltage felt by the artificial atom (flux qubit) locating at the end of the transmission line ($x=0$) is 
$\frac{1}{2}\mathcal{V}_\mathrm{p}e^{-\mathrm{i}\nu t}+\frac{1}{2}\mathcal{V}_\mathrm{c}e^{-\mathrm{i}\omega t}+\mathrm{c}.\mathrm{c}.$ Note that the probe (control) field with frequency $\nu$ ($\omega$) is nearly resonant with the transition channel $|1\rangle\leftrightarrow|3\rangle$ ($|2\rangle\leftrightarrow|3\rangle$). Thus under rotation wave approximation, the Hamiltonian of artificial atom can be written as
\begin{eqnarray}
\hat{H}_{\mathrm{s}}&=&\sum_{i=1}^{3}{\hbar\omega_{i}}|i\rangle\langle i|
\nonumber
\\
&&-\frac{\hbar}{2}\left(\Omega_\mathrm{p}e^{-\mathrm{i}\nu t}|3\rangle\langle1|+\Omega e^{-\mathrm{i}\omega t}|3\rangle\langle2|+\mathrm{H.c.}\right),
\label{SemiHamiltonian}
\end{eqnarray}
where the Rabi frequencies are defined as $\Omega_{\mathrm{p}}=\frac{1}{\hbar}|q_{31}|\mathcal{V}_{\mathrm{p}}$ and $\Omega=\frac{1}{\hbar}|q_{32}|\mathcal{V}_{\mathrm{c}}$, respectively. Here $q_{ij}$ is matrix elements of effective charge operator $\hat{q}=\frac{2e \beta}{1+2 \alpha+2 \beta}{\hat{n}_{\mathrm{m}}}$ (like dipole moment matrix elements in atomic physics).

It is convenient to work in the interaction picture. Here we define 
\begin{equation}
\hat{H}_{\mathrm{0}}=\hbar\omega_{1}|1\rangle\langle 1|+\hbar(\omega_{1}+\nu)|3\rangle\langle 3|+\hbar(\omega_{1}+\nu')|2\rangle\langle 2|
\end{equation}
with $\nu'=\nu-\omega$. After using $\hat{H}_{\mathrm{s}}\rightarrow e^{\mathrm{i}\hat{H}_0 t}\hat{H}_{\mathrm{s}}e^{-\mathrm{i}\hat{H}_0 t}$, we find in the interaction picture, the Hamiltonian of this driven three-level system reads
\begin{eqnarray}
\hat{H}_{\mathrm{s}}&=&-\hbar\Delta_{\mathrm{p}}|3\rangle\langle3|-\hbar\left(\Delta_{\mathrm{p}}-\Delta\right)|2\rangle\langle2|
\nonumber
\\
&&-\frac{\hbar}{2}\left(\Omega_\mathrm{p}|3\rangle\langle1|+\Omega|3\rangle\langle2|+\mathrm{H.c.}\right),
\label{SemiHamiltonianInt}
\end{eqnarray}
where the detunings are defined as $\Delta_{\mathrm{p}}=\nu-(\omega_{3}-\omega_{1})$, $\Delta=\omega-(\omega_{2}-\omega_{1})$. The dynamics of the atom can be described by the master equation
\begin{equation}
\frac{\mathrm{d}\hat{\rho}}{\mathrm{d}t}=\frac{1}{\mathrm{i}\hbar}\left[\hat{H}_{\mathrm{s}},~\hat{\rho}\right]+\mathcal{L}\left[\hat{\rho}\right],
\label{MasterEquation}
\end{equation}
where the Lindblad term is defined by
\begin{eqnarray}
\mathcal{L}\left[\hat{\rho}\right]&=&\Gamma_{31}\left(|1\rangle\langle1|-|3\rangle\langle3|\right)+\Gamma_{32}\left(|2\rangle\langle2|-|3\rangle\langle3|\right)
\nonumber
\\
&&+\Gamma_{21}\left(|1\rangle\langle1|-|2\rangle\langle2|\right)-\sum_{i\neq j}\gamma_{ij}\rho_{ij}|i\rangle\langle j|.
\label{Loperator}
\end{eqnarray}
Here $\Gamma_{ij}$ ($i > j$) are the relaxation rates between the levels $|i\rangle$ and $|j\rangle$, as defined in the main text. $\gamma_{ij} = \gamma_{ji}$ are the damping rates of the off-diagonal terms. Specifically,
$\gamma_{12}=\Gamma_{21}/2+\gamma_{12}^{\varphi}/2$,
$\gamma_{13}=(\Gamma_{32}+\Gamma_{31})/2+\gamma_{13}^{\varphi}/2$,
$\gamma_{23}=(\Gamma_{32}+\Gamma_{31}+
\Gamma_{21})/2+\gamma_{23}^{\varphi}/2$, with $\gamma_{ij}^{\varphi}$
being the pure dephasing. The pure dephasing $\gamma_{ij}^{\varphi}$ and the photon loss rate $\gamma_{i}$ defined in main text have relations: $\gamma_{12}^{\varphi}=\gamma_{2}$, $\gamma_{13}^{\varphi}=\gamma_{3}$, $\gamma_{23}^{\varphi}=\gamma_{2}+\gamma_{3}$.

The external fields can induce an effective charge $\hat{q}$ on the artificial atom at $x=0$ (playing
a role of atomic polarization). The corresponding expectation value is $q_{\mathrm{e}}=\mathrm{Tr}{[\hat{q}\hat{\rho}]}$. Here we are interest in the components at frequencies $\nu$ and $\nu'$, which can be written as
\begin{equation}
q_{13}\rho_{31}e^{-\mathrm{i}\nu t}+q_{12}\rho_{21}e^{-\mathrm{i}\nu' t}+{\mathrm{c}}.{\mathrm{c}}.
\end{equation}
Without loss of generality, we can replace $q_{ij}$ by its absolute value $|q_{ij}|$ (i.e., move the phase factor of $q_{ij}$ into $\rho_{ji}$). This oscillating charge as a point-like source can re-emitting microwave into the transmission line.
The net wave in the transmission line should be superposition of the externally applied waves and the re-emitted waves by the artificial atom, which satisfies the relevant one dimensional wave equation:
\begin{equation}
\partial_{xx}V(x,t)-v_{\mathrm{g}}^{-2}\partial_{tt}V(x,t)=l\delta(x)\partial_{tt}q_{\mathrm{e}}(t),
\label{WaveEQ}
\end{equation}
where $v_{\mathrm{g}}=1/\sqrt{lc}$ is the group velocity, $l$ and $c$ are characteristic inductance and capacitance per unit length. Specifically, the components of the field oscillating at frequencies $\nu$ and $\nu'$ can be written as
\begin{eqnarray}
V(x,t)&=&\frac{1}{4}\mathcal{V}_{\mathrm{p}}e^{\mathrm{i}kx-\mathrm{i}\nu t}\theta(-x)+\frac{1}{4}T_{a}\mathcal{V}_{\mathrm{p}}e^{-\mathrm{i}kx-\mathrm{i}\nu t}\theta(-x)
\nonumber
\\
&&+\frac{1}{4}\sqrt{\frac{\nu'}{\nu}}T_{b}\mathcal{V}_{\mathrm{p}}e^{-\mathrm{i}k'x-\mathrm{i}\nu' t}\theta(-x)+\mathrm{c}.\mathrm{c}.
\label{netwave}
\end{eqnarray}
where $T_a$ and $T_b$ are defined as photon number transmission coefficients. Substituting Eqs.~\eqref{netwave} into Eqs.~\eqref{WaveEQ}, and after some calculations, we have
\begin{eqnarray}
T_a&=&1+2\mathrm{i}\frac{\Gamma_{31}}{\Omega_{\mathrm{p}}}\rho_{31}
\label{TTa}
\\
T_b&=&2\mathrm{i}\frac{\sqrt{\Gamma_{31}\Gamma_{21}}}{\Omega_{\mathrm{p}}}\rho_{21}
\label{TTb}
\end{eqnarray}

To give the single-photon condition of the input field, we can define an average number of photons per interaction time $2\pi/\Gamma_{31}$ as $N=2{\pi}P/(\hbar\nu\Gamma_{31})$. Here the probe power is $P=(\mathcal{V}_{\mathrm{p}}^{\left(\mathrm{in}\right)})^2/(2Z)$, where $\mathcal{V}_{\mathrm{p}}^{(\mathrm{in})}=\mathcal{V}_{\mathrm{p}}/2$ is the complex amplitude of input field. By using $\Omega_{\mathrm{p}}=\frac{1}{\hbar}|q_{31}|\mathcal{V}_{\mathrm{p}}$ and $\Gamma_{31}=\frac{2} {\hbar} Z \omega_{31} {\left\vert q_{31}\right\vert}^{2}$, one can rewritten the average number of photons as $N=\pi\Omega_{\mathrm{p}}^2/(2\Gamma_{31}^2)$. Note that the single-photon condition is $N\ll 1$, thus it can be equally written as  $\Omega_{\mathrm{p}}\ll\Gamma_{31}$. To generate frequency down-converted photons with high effeciency, the average number of input photons within the time scale $2\pi/\Gamma_{21}$ should also much less than $1$, resulting in the condition $\Omega_{\mathrm{p}}\ll\Gamma_{21}$. If the Rabi frequency of control field satisfies Eq.~\eqref{opt-condition-DC}, we find $\Omega\simeq\sqrt{\Gamma_{31}\Gamma_{21}}\geq\min(\Gamma_{31},\Gamma_{21})$. Thus if the input probe $\Omega_{\mathrm{p}}$ satisfies the conditions $\Omega_{\mathrm{p}}\ll\Gamma_{31}$ and $\Omega_{\mathrm{p}}\ll\Gamma_{21}$, we furthermore have $\Omega_{\mathrm{p}}\ll\Omega$.

If the input field $\Omega_{\mathrm{p}}$ is sufficiently weak ($\Omega_{\mathrm{p}}\ll\Omega$, $\Omega_{\mathrm{p}}\ll\Gamma_{31}$, $\Omega_{\mathrm{p}}\ll\Gamma_{21}$), the analytical expressions of the induced coherence between levels $|1\rangle$ and $|2\rangle$ ($|3\rangle$), up to the first order of $\Omega_{\mathrm{p}}$, can be written as
\begin{eqnarray}
\rho_{21}^{(1)}=\frac{-\frac{1}{4}\Omega_{\mathrm{p}}\Omega}{\left[\mathrm{i}\left(\Delta_{\mathrm{p}}-\Delta\right)
-\gamma_{12}\right]\left(\mathrm{i}\Delta_{\mathrm{p}}-\gamma_{13}\right)+\frac{\Omega^2}{4}},
\label{rho21}
\\
\rho_{31}^{(1)}=\frac{-\mathrm{i}\frac{\Omega_{\mathrm{p}}}{2}\left[\mathrm{i}\left(\Delta_{\mathrm{p}}-\Delta\right)
-\gamma_{12}\right]}{\left[\mathrm{i}\left(\Delta_{\mathrm{p}}-\Delta\right)
-\gamma_{12}\right]\left(\mathrm{i}\Delta_{\mathrm{p}}-\gamma_{13}\right)+\frac{\Omega^2}{4}}.
\label{rho31}
\end{eqnarray}
Substituting above results into Eqs.~\eqref{TTa} and \eqref{TTb}, we find Eqs.~\eqref{TranTa} and \eqref{TranTb} in the main text (which obtained by treating the input field as single photon) can be repeated. This means that in this weak-inout (or single-photon) limit, the two methods are equivalent.  

When the power of the incident microwave is increased ($\Omega_\mathrm{p}\sim\Gamma_{31}$ or $\Omega_\mathrm{p}>\Gamma_{31}$), the atom will be saturated. In this nonlinear region, the analytical calculations by taking weak-input-field (or single-photon) limit fail. But we can still numerically calculate the steady-state density matrix elements $\rho_{ij}$ by utilizing the mater equation \eqref{MasterEquation}, and further obtain transmission coefficients through Eqs.~\eqref{TTa} and \eqref{TTb}. The results are shown in Figs.~\ref{nonlinear}(a)-(c) and the corresponding discussions are given in Sec.~\ref{Numerical} in the main text.


\end{document}